\pdfoutput=1

\documentclass[11pt]{article}

\usepackage[final]{acl}

\usepackage{times}
\usepackage{latexsym}

\usepackage[T1]{fontenc}

\usepackage[utf8]{inputenc}

\usepackage{microtype}

\usepackage{inconsolata}

\usepackage{hyperref}
\usepackage{adjustbox}
\usepackage{url}
\usepackage{subcaption}
\usepackage{booktabs}
\usepackage{multirow}
\usepackage{xcolor}
\usepackage{marginnote}
\usepackage{amsfonts}
\usepackage{graphicx}
\usepackage{wrapfig}
\usepackage{lipsum}
\makeatletter
\renewcommand{\@fnsymbol}[1]{\dag} 
\makeatother
\title{Advancing Test-Time Adaptation in Wild Acoustic Test Settings}

\author{
 \textbf{Hongfu Liu},
 \textbf{Hengguan Huang}\thanks{Corresponding Author},
 \textbf{Ye Wang}
\\
 \textsuperscript{1}National University of Singapore \\
 \texttt{\{hongfu,wangye\}@comp.nus.edu.sg, huang.hengguan@u.nus.edu}
}

\begin{document}
\maketitle
\begin{abstract}

Acoustic foundation models, fine-tuned for Automatic Speech Recognition (ASR), suffer from performance degradation in wild acoustic test settings when deployed in real-world scenarios. Stabilizing online Test-Time Adaptation (TTA) under these conditions remains an open and unexplored question. Existing wild vision TTA methods often fail to handle speech data effectively due to the unique characteristics of high-entropy speech frames, which are unreliably filtered out even when containing crucial semantic content. Furthermore, unlike static vision data, speech signals follow short-term consistency, requiring specialized adaptation strategies. In this work, we propose a novel wild acoustic TTA method tailored for ASR fine-tuned acoustic foundation models. Our method, Confidence-Enhanced Adaptation, performs frame-level adaptation using a confidence-aware weight scheme to avoid filtering out essential information in high-entropy frames. Additionally, we apply consistency regularization during test-time optimization to leverage the inherent short-term consistency of speech signals. Our experiments on both synthetic and real-world datasets demonstrate that our approach outperforms existing baselines under various wild acoustic test settings, including Gaussian noise, environmental sounds, accent variations, and sung speech~\footnote{Code is publicly available at \href{https://github.com/Waffle-Liu/CEA}{https://github.com/Waffle-Liu/CEA}.}.


\end{abstract}

\section{Introduction}

\begin{figure}
    \centering
    \includegraphics[scale=0.5]{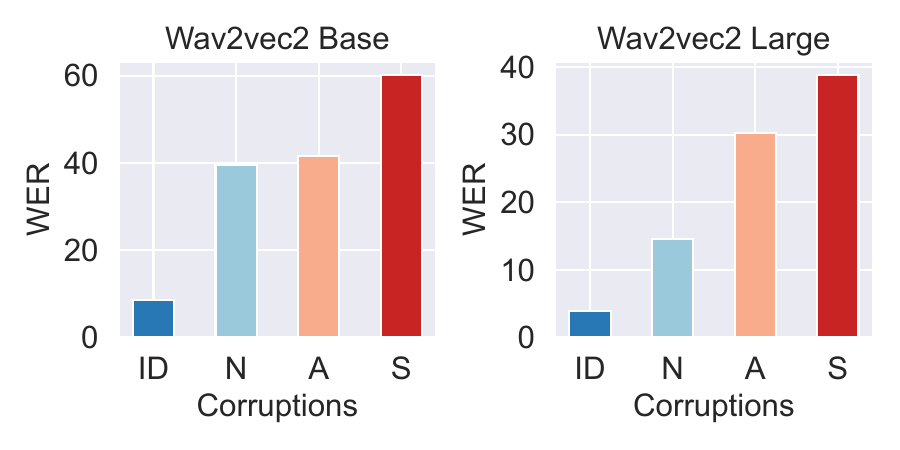}
    \caption{Robustness analysis of Wav2vec2 Base and Large under wild acoustic test settings including 1) Noise (\textbf{N}): additive noises on LibriSpeech test-other set, 2) Accent (\textbf{A}): accents of L2 learners on L2-Arctic subset 3) Singing (\textbf{S}): sung speech on DSing test set. In-Domain (\textbf{ID}) indicates the performance on LibriSpeech test-other set without additive noises. WER is short for Word Error Rate. }
    \label{fig:robustness}
    \vspace{-2mm}
\end{figure}

Deep learning-based acoustic models have exhibited remarkable performance in scenarios where the training and test sets adhere to the independent and identically distributed (i.i.d) assumption. However, real-world applications frequently involve domain shifts between training and test sets, such as noise variations due to environmental sounds~\citep{noisyenv}, and timbre variations due to accent or pronunciation changes~\citep{reproadapter}. While recent acoustic foundation models, such as Wav2vec2~\citep{wav2vec2}, fine-tuned on Automatic Speech Recognition (ASR) achieve excellent performances, they exhibit notable performance degradation when confronted with the test-time speech in the wild, as depicted in Figure~\ref{fig:robustness}. Consequently, there exists an emergent demand to adapt these acoustic foundation models in wild acoustic test settings when deployed in the real world.  

Prior methods for mitigating domain shifts require access to domain-specific source data under the unsupervised domain adaptation setting~\cite{adaptation_overview}, limiting the application to online scenarios where speech data come from the wild world with mixed distribution shifts. Test-Time Adaptation (TTA) emerges as a critical paradigm for addressing distribution shifts at inference time, enabling online updates of models on test data in a source-free way. Recent work, SUTA~\citep{suta}, presents a pilot study on TTA for ASR models by applying entropy minimization to speech frame adaptation, demonstrating impressive performance on single-utterance TTA. However, SUTA focuses on mild test settings, \textit{e.g.}, testing on speech with synthetic and real noises. In the dynamic wild world, acoustic foundation models may face arbitrary test speech data with severe distribution shifts, such as sung speech. As such, stabilizing online TTA under wild acoustic test settings remains an open and unexplored question. Recent work, SAR~\citep{sar}, proposes an efficient optimization scheme for stabling online TTA in the wild vision test settings. However, direct adoption of SAR to speech data is challenging because SAR characterizes high-entropy noisy speech samples as unreliable and potentially harmful for model adaptation and proposes to filter them out for stabling under wild vision test settings.

In this work, we empirically identify a substantial proportion of noisy frames within non-silent speech segments under wild acoustic test settings. We observe that these frames contain vital semantic information crucial for accurate recognition and merely discarding these noisy frames may adversely affect model performance. Consequently, rather than excluding these noisy non-silent frames, we propose Confidence Enhanced Adaptation (CEA), which performs frame-level adaptation using a confidence-aware weight scheme. CEA prioritizes uncertain frames and encourages models to focus more on these uncertain frames by `denoising' their intermediate representations. Additionally, we emphasize that frames within a short speech segment are temporally coherent, largely due to the consistent nature of phonemic content within such windows, thus proposing short-term consistency regularization to stabilize wild acoustic TTA. This contrasts with image samples in a batch, which are frequently treated as independent entities. We conduct a wide range of experiments for ASR fine-tuned acoustic foundation models on both synthetic and real-world datasets, systematically assessing the model's robustness against Gaussian noises, environmental sounds, accents of second language (L2) learners, and singing (a.k.a. sung speech). The experimental results demonstrate the effectiveness of our method under wild acoustic test settings.

In summary, our contributions are summarized as follows:
\begin{itemize}
    \item We are the first to address wild acoustic TTA and observe that in wild acoustic test settings high-entropy noisy speech frames are often located within non-silent segments crucial for semantic understanding. We introduce CEA with a confidence-aware weight scheme to efficiently adapt noisy non-silent frames.
    \item We highlight the consistent nature of phonemic content within short speech segments and introduce short-term consistency regularization to further stabilize acoustic wild TTA.
    \item We perform a wide range of experiments on both synthetic and real-world datasets, including new experiments on real-world sung speech datasets for the first time. Empirical results substantiate the efficacy of our method under wild acoustic test settings.
\end{itemize}

\section{Related Work}
\label{related}

\subsection{Test-Time Adaptation. } Test-time adaption plays an essential role in addressing distribution shifts encountered in test samples, enabling online updates of models during the test phase using unsupervised objectives. Most prior TTA methods in the computer vision domain rely on Batch Normalization layers~\citep{bn,ttn,eata} and assume sample independence within the same batch~\citep{cota,note} despite addressing non-i.i.d data streams in fluctuating environments, rendering them less applicable to speech data. Additionally, real-world data shifts, including both covariate and label shifts, pose significant challenges for deployment~\citep{wilds,sar,ods}. From another line of research, \cite{huang2022extrapolative} introduced a training-free TTA framework that handles non-stationary covariate shifts by leveraging a latent continuous-time dynamical system to infer model parameters. Recent work provides a pilot study on TTA for ASR models under mild test settings~\citep{suta}, and improves TTA for general ASR models via sequence-level generalized entropy minimization~\citep{suta}. Our work focuses more on stabilizing online TTA for ASR models under wild acoustic settings. We empirically analyze frame-level entropy distribution and underscore the short-term consistency nature of speech signals.

\subsection{Robustness for ASR. } There is a long history of developing robust speech recognition methods~\cite{speechrob2}. For example, \citet{huang2017improve,huang2018wavenet} enhances the robustness of acoustic models by incorporating higher-order features, while \citet{huang2019recurrent,huang2021strode} improves the noise robustness by guiding the model to focus on inferred informative latent acoustic events. Different from improving model robustness by training with large-scale augmented data~\citep{whisper}, there are various adaptation approaches for acoustic distribution shifts.  Recent works explore input reprogramming~\citep{reprogramming,reproxl} with supervised optimization targets. Unsupervised domain adaptation (UDA) approaches investigate the feature alignment~\citep{feature_align}, data augmentation~\citep{dataaug}, domain adversarial training~\citep{adver1,adver2}, knowledge distillation~\citep{distill}, and self-training~\cite{distill}. However, these methods require access to the source data with severe latency and heavy computation, and tackle distinct acoustic shifts, such as speaker~\citep{speaker2} and accent adaptation~\citep{reproadapter} in isolation, limiting their applications to online scenarios. Early test-time method for traditional acoustic models, LUHC, with parameterized activation functions~\citep{speakadapt,unsupadapt} also deals with specific acoustic shifts, lacking the generalization ability under wild acoustic test settings. Despite the success of prior adaptation methods, the development of online TTA for modern ASR-fined acoustic foundation models under wild acoustic test settings remains an open and unexplored question.

\section{Preliminary}
We center our focus on the fully Test-Time Adaptation framework, characterized by episodic model adaptation, where the model is reset after processing each utterance. We denote the ASR fine-tuned acoustic foundation model as $f_{\Theta}(y | x)$. We investigate the popular acoustic foundation models such as Wav2vec2~\citep{wav2vec2}, HuBERT~\citep{hubert}, WavLM~\citep{wavlm}, which can be typically decomposed into two constituent components: a feature extractor $g_{\phi}(z|x)$, parameterized by $\phi$, and a transformer encoder $h_{\theta}(y|z)$, parameterized by $\theta$. This decomposition is expressed as:
\begin{equation}
    f_{\Theta}(y|x) = h_\theta(g_\phi(x))
\end{equation}
where $\Theta = \{\theta, \phi\}$ represents the collective set of model parameters. The feature extractor $g_\phi$ takes as input waveform audio or log-mel spectrogram. The transformer encoder $h_\theta$ serves as an audio encoder and outputs acoustic representations. Considering a test-time speech sequence $x_{1:n}$ of variable length $n$ in the wild, typically with arbitrary domain shifts, the primary objective entails adapting the acoustic foundation model $f_\Theta$ to enhance its performance for $x_{1:n}$.

\section{Method}

\label{method}
\begin{figure*}
    \centering
    \includegraphics[scale=0.37]{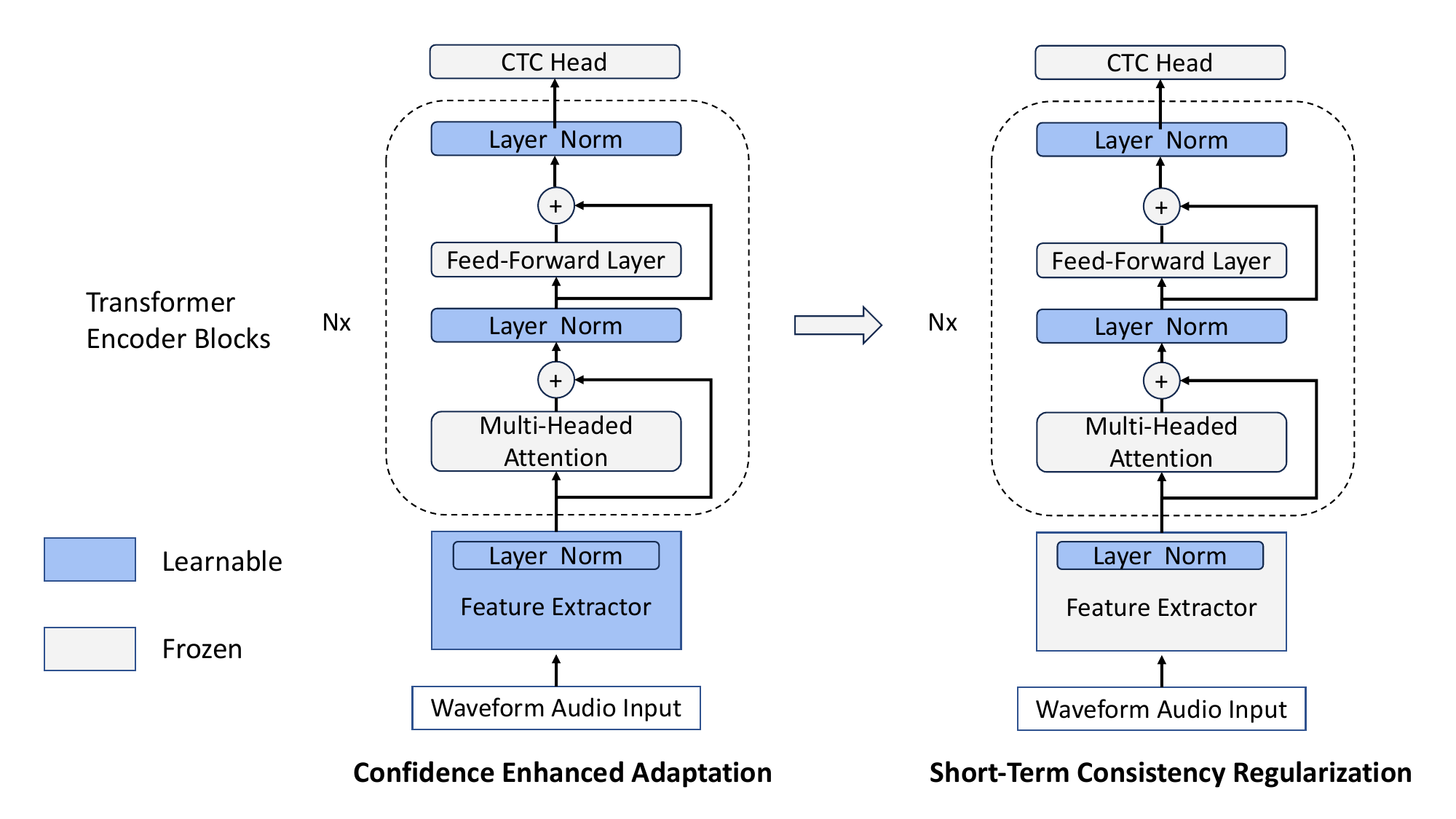}
    \caption{The overall framework of the proposed method. The figure takes a Connectionist Temporal Classification (CTC) based acoustic foundation model as an example. This framework involves two steps. The confidence enhanced adaptation is first performed to boost the reliability of noisy frames. The temporal consistency regularization is employed across the entire input sequence and jointly optimized with entropy minimization.}
    \label{fig:main}
\end{figure*}

In this section, we first analyze the common source of domain shifts in the wild acoustic test settings, and then provide our findings and methods for addressing the wild acoustic shifts. The overview of our method is presented in Figure \ref{fig:main}. 

\subsection{Wild Acoustic Test Settings}

Wild acoustic distribution shifts encountered within the speech domain may originate from several sources, including: 

\textbf{Speaker Changes}. Timbre variations in speech stemming from changes in the speaker's identity.

\textbf{Environmental Noises}. Perturbations introduced by ambient noises in the recording environments.

\textbf{Pronunciation Changes}. Alteration in pronunciation characteristics such as accent or singing.

\textbf{Text-Domain Changes}. Shifts in the linguistic content or context of the speech data.

It is noteworthy that speaker changes, environmental noises, and pronunciation changes are typically categorized as covariate shift, as they pertain to variations in the input data distribution. In contrast, text-domain changes are categorized as label shift, as they involve alterations in the output distribution. Furthermore, it is important to acknowledge that real-world speech data often exhibit shifts stemming from multiple sources simultaneously, rendering the adaptation under wild acoustic test settings complex and challenging.

\begin{figure}[h]
    \centering
    \includegraphics[scale=0.37]{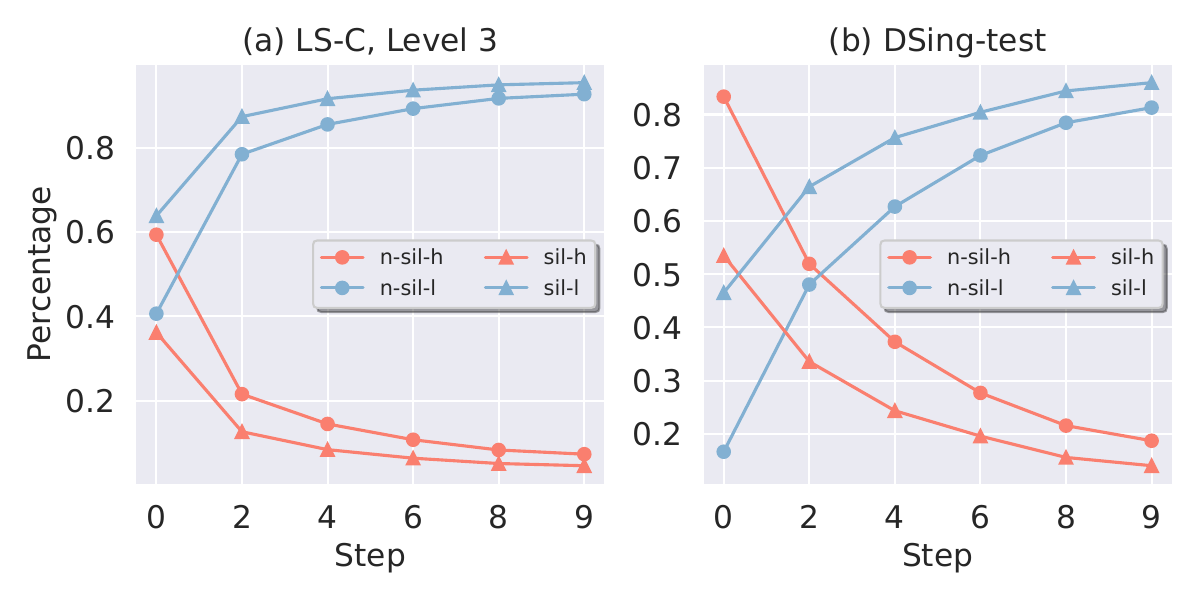}
    \caption{Frame-Level Entropy Distribution in ASR fine-tuned Acoustic Foundation Models: the entropy distributions are computed for Wav2vec2 Base models on the LibriSpeech noise-corrupted test-other and DSing test datasets across adaptation steps. We employ a threshold of $0.4*\ln{C}$, as recommended in \citet{eata}, where $C$ represents the number of task classes. Frames with entropy values exceeding this threshold are highlighted in red, indicating high-entropy (h) frames, while low-entropy (l) frames are marked in blue. We use $\bullet$ to denote non-silent (non-sil) frames and $\triangle$ for silent (sil) frames and take the blank symbol as an approximate indicator. The training steps range from 0 to 9, and the results presented in each subfigure are based on the average of 100 random samples.}
    \label{fig:entropy}
    \vspace{-5mm}
\end{figure}

\subsection{Confidence Enhanced Adaptation}

To gain insights into the behavior of ASR fine-tuned acoustic foundation models under wild acoustic test settings, we empirically analyze the frame-level entropy distribution of speech data in the wild. We conducted experiments using both the LibriSpeech test-other dataset, which was deliberately corrupted by additive Gaussian noises, and the sung speech dataset, DSing-test. These experiments were performed with the ASR fine-tuned Wav2vec2 Base model. We subsequently evaluated the percentages of high-entropy and low-entropy frames for both non-silent and silent speech segments. The classification of frames as silent or non-silent was determined based on pseudo labels derived from model predictions.

As illustrated in Figure \ref{fig:entropy}, our findings reveal that, prior to any adaptation (Step=0), within the non-silent frames category, there exists a prevalence of high-entropy frames compared to low-entropy ones for Base models. Conversely, the opposite trend is observed within the silent frames category. It is worth noting that existing literature~\citep{sar} provides heuristic insights suggesting that high-entropy samples may be unreliable and could potentially have a detrimental impact on model adaptation. However, it is crucial to recognize that these noisy frames contain essential content information that is critical for speech recognition. While prior research suggests that filtering out such unreliable samples may aid in stabilizing adaptation under wild vision test settings and improving performance, this approach proves infeasible in our specific case. 


In response, rather than dropping these high-entropy noisy frames, we propose a learning-based approach, Confidence Enhanced Adaptation (CEA), which performs frame-level adaptation using a confidence-aware weight scheme. CEA prioritizes uncertain frames and encourages models to focus more on these uncertain frames by ‘denoising’ their intermediate representations. Denoting $\hat{y}^c_{i} = f_{\Theta}(c | x_{1:n} )$ as the predicted probability of class $c$ for $i$-th frame, we quantify uncertainty through entropy, defined as:

\begin{equation}
E(x_{i}) = - \sum_c \hat{y}^c_{i} \log \hat{y}^c_{i}
\end{equation}

Instead of heuristically relying on manually set thresholds for filtering out data samples of high entropy, CEA utilizes pseudo labels $\hat{y}_i$ assigned to each frame $x_i$ and applies entropy minimization with a confidence-aware weight scheme on these non-silent noisy frames, without the need for setting thresholds. Specifically, we define the confidence-aware optmization scheme as follows:

\begin{equation}
  \mathop{\min}_{\Theta^\prime = \{\phi,\theta_{LN}\}} \sum^n_{i=1} S(x_{i})E(x_{i})
\end{equation}

where $\theta_{LN}$ denotes the affine parameters associated with layer normalization in the transformer encoder $h$, and $S(x_{i})$ represents confidence-aware frame-level weights, defined as:

\begin{equation}
S(x_{i}) = \frac{1}{1+\exp(-E(x_{i}))} \mathbb{I}_{\hat{y}i \neq c_0}(x_{i})
\end{equation}

where $c_0$ signifies the index corresponding to silent frames, and $\mathbb{I}$ is an indicator function.  Such design empowers the model to assign greater importance to frames where it exhibits lower confidence. The increased weight encourages the model to focus more on these uncertain frames during adaptation, potentially leading to heightened model confidence on such frames. Note that this adaptation process entails an update of the feature extractor $g_\phi$. This empowers models with the capability to adapt to wild acoustic shifts, even in the presence of substantial covariate shifts. As evidenced in Figure~\ref{fig:entropy}, the count of high-entropy frames diminishes while low-entropy frame counts increase with each adaptation step, underscoring the effectiveness of CEA.

\subsection{Short-Term Consistency Regularization}

In the domain of speech signal processing, a salient characteristic is the short-term stability, where successive speech frames often convey the same phoneme or speech unit. This intrinsic temporal correlation is a defining attribute of speech data, making it essential for stabilizing online TTA under wild acoustic test settings. Nevertheless, prior TTA methods largely overlook this inherent temporal correlation within individual speech sequences.

To address this limitation, we propose a feature-wise short-term consistency regularization technique. We perform this regularization step after the confidence enhanced adaptation process. This sequencing is deliberate as introducing temporal regularization over representations of noisy frames can potentially confuse models and yield undesirable optimization outcomes. Concretely, the regularization is jointly optimized alongside entropy minimization, as represented by the following equation:

\begin{equation}
\mathop{\min}_{\Theta_{LN}} \sum^n_{i=1} E(x_{i}) + \alpha\sum^{n-k+1}_{i=1}\!|| z^\prime_{k+i-1} -z^\prime_{i} ||_{2} \mathbb{I}_{\hat{y}_i \neq c_0}(x_{i})
\end{equation}

where $\alpha$ denotes the weight assigned to the regularization loss, and $\Theta_{LN}$ represents the affine parameters associated with layer normalization across the entire acoustic foundation model. Here, $z_{i}$ signifies the feature representation of $i$-th frame obtained from the fine-tuned feature extractor, and $z^\prime_{i}$ represents the modified feature representation achieved through a parameter-free self-attention operation. The parameter $k$ denotes the size of the window considered as the neighborhood of frame $x_{i}$. This regularization technique effectively captures the inherent temporal consistency found in speech data by compelling the representation of $x_{i}$ to closely resemble that of its neighboring frames within a predefined window. Despite the possible peaky behavior of CTC, the proposed temporal consistency can be treated as introducing the inductive bias of "short-term stability" in the adaptation~\citep{rabiner2007introduction}.

\section{Experiments}
\begin{table*}[htbp]
\begin{center}

\begin{tabular}{cccccccccccccccccccccccccccccccccccccccccccccccc}
\toprule[1pt]
 \multicolumn{6}{c}{Method} & \multicolumn{6}{c}{Level 0} & \multicolumn{6}{c}{Level 1} & \multicolumn{6}{c}{Level 2} & \multicolumn{6}{c}{Level 3} & \multicolumn{6}{c}{Level 4} & \multicolumn{6}{c}{Level 5} & \multicolumn{6}{c}{Average} \\ 
\midrule[0.5pt]
  \multicolumn{6}{c}{$\delta$} & \multicolumn{6}{c}{0} & \multicolumn{6}{c}{0.005} & \multicolumn{6}{c}{0.01} & \multicolumn{6}{c}{0.015} & \multicolumn{6}{c}{0.02} & \multicolumn{6}{c}{0.03} & \multicolumn{6}{c}{} \\
\midrule[0.5pt]
 \multicolumn{6}{c}{Source} & \multicolumn{6}{c}{$8.6$} & \multicolumn{6}{c}{$13.9$} & \multicolumn{6}{c}{$24.4$} & \multicolumn{6}{c}{$39.5$} & \multicolumn{6}{c}{$54.5$} & \multicolumn{6}{c}{$75.7$} & \multicolumn{6}{c}{$31.6$}  \\ 
\multicolumn{6}{c}{Tent} & \multicolumn{6}{c}{$7.7$} & \multicolumn{6}{c}{$11.6$} & \multicolumn{6}{c}{$19.7$} & \multicolumn{6}{c}{$32.2$} & \multicolumn{6}{c}{$46.3$} & \multicolumn{6}{c}{$69.2$} & \multicolumn{6}{c}{$31.1$}  \\ 
\multicolumn{6}{c}{SAR} & \multicolumn{6}{c}{$8.2$} & \multicolumn{6}{c}{$12.7$} & \multicolumn{6}{c}{$21.5$} & \multicolumn{6}{c}{$35.0$} & \multicolumn{6}{c}{$49.2$} & \multicolumn{6}{c}{$72.0$} & \multicolumn{6}{c}{$33.1$}  \\ 
\multicolumn{6}{c}{TeCo} & \multicolumn{6}{c}{$7.6$} & \multicolumn{6}{c}{$13.6$} & \multicolumn{6}{c}{$19.7$} & \multicolumn{6}{c}{$32.2$} & \multicolumn{6}{c}{$46.3$} & \multicolumn{6}{c}{$69.3$} & \multicolumn{6}{c}{$31.5$}  \\ 
 \multicolumn{6}{c}{SUTA} & \multicolumn{6}{c}{$\textbf{7.3}$} & \multicolumn{6}{c}{$10.9$} & \multicolumn{6}{c}{$16.7$} & \multicolumn{6}{c}{$24.6$} & \multicolumn{6}{c}{$34.7$} & \multicolumn{6}{c}{$\textbf{56.5}$} & \multicolumn{6}{c}{$25.1$}  \\ 
 \multicolumn{6}{c}{Ours} & \multicolumn{6}{c}{$\textbf{7.3}$} & \multicolumn{6}{c}{$\textbf{10.7}$} & \multicolumn{6}{c}{$\textbf{16.2}$} & \multicolumn{6}{c}{$\textbf{24.0}$} & \multicolumn{6}{c}{$\textbf{34.1}$} & \multicolumn{6}{c}{$\textbf{56.5}$} & \multicolumn{6}{c}{$\textbf{24.8}$}  \\  

\bottomrule[1pt]
\end{tabular}
\caption{WER (\%) results on LS-C over five severity levels $\delta$ of Gaussian noises using Wav2vec2 Base with greedy decoding. $\delta = 0 $ represents the uncorrupted case. The best results are bold.}
\label{table:ls}
\vspace{-3mm}
\end{center}
\end{table*}



In this section, we undertake an evaluation of the robustness of ASR fine-tuned acoustic foundation models under wild acoustic test settings. We discuss the robustness against synthetic noises including Gaussian noises and real-world environmental sounds in Section~\ref{noise}, real-world data shifts including L2 accents and singing voice (sung speech) in Section~\ref{realworld}, and decoding strategy pertaining to language models in Section~\ref{decoding}. We provide more evaluation results using various acoustic models in Appendix~\ref{app: more_models}.

\label{experiments}

\subsection{Experimental Setup}

\paragraph{Datasets. }Our experiments involve the utilization of four distinct datasets: two synthetic and two real-world datasets. The first synthetic dataset, named LS-C, represents the LibriSpeech~\citep{librispeech} test-other set Corrupted by additive Gaussian noises. We introduce five levels of severity to simulate various degrees of corruption as per ~\cite{benchmarking} for evaluating the trend of model robustness. Higher levels indicate more severe corruption although heavily corrupted speech data may not be common cases in the real world. Subsequently, the second synthetic dataset, named LS-P, is the LibriSpeech test-other set Perturbed by real-world environmental sounds. This dataset encompasses eight diverse types of environmental sound, including Air Conditioner, Babble, Munching, Shutting Door, Vacuum Cleaner, Airport Announcements, Copy Machine, and Typing. These environmental sounds are from the MS-SNSD noise test set~\citep{noisyenv}. Each type is added to the original audio with five distinctive signal-to-noise ratios (SNRs) representing five levels of severity. Our study further extends to two real-world datasets. The L2-Arctic~\citep{l2arctic} dataset comprises speech data from second language (L2) learners originating from six countries with different first languages (L1): Arabic, Mandarin, Hindi, Korean, Spanish, and Vietnamese. Furthermore, we broaden our investigation to encompass music datasets, DSing~\citep{dsing} and Hansen \citep{hansen}, featuring singing voice (sung speech). More details of dataset statistics can be found in Appendix~\ref{app:data} and details of implementation can be found in Appendix~\ref{app:implement}.  

\paragraph{Baselines. }To assess the adaptation performance of our proposed method, we consider the following TTA baselines. \textbf{Tent}~\citep{tent} adapt transformation layers with the objective of entropy minimization. Despite it being initially proposed for batch normalization, we refer to updating the affine parameters of layer normalization as Tent in our work. In addition, we involve the baseline \textbf{TeCo}~\citep{teco}, originally proposed for video classification with temporal coherence regularization, due to its applicability to sequential data. Our comparison also includes the \textbf{SAR}~\citep{sar}, specifically designed to address data shifts in the dynamic wild world. Furthermore, we also introduce comparisons with \textbf{SUTA}~\citep{suta} using entropy minimization and minimum class confusion, and \textbf{SGEM}~\citep{sgem} using sequential-level generalized entropy minimization in conjunction with beam search employing language models. 

\begin{table}[h]
\centering
\begin{minipage}{0.47\textwidth}
\centering
\begin{tabular}{cccccc}
    \toprule
      & 10    & 5     & 0     & -5    & -10 \\
    \midrule
    Source & 28.1  & 43.9  & 65.0  & 83.4  & 94.2 \\
    Tent  & 22.6  & 36.1  & 56.6  & 77.9  & 91.4 \\
    SAR   & 24.5  & 39.1  & 59.9  & 79.9  & 92.1 \\
    TeCo  & 22.5  & 36.2  & 56.6  & 77.9  & 91.3 \\
    SUTA  & 17.7  & 26.1  & 41.2  & 62.7  & 82.7 \\
    Ours  & \textbf{17.5}  & \textbf{25.6}  & \textbf{40.6}  & \textbf{61.6}  & \textbf{82.2} \\
    \bottomrule
    \end{tabular}%
\caption{WER (\%) results on \textbf{Air Conditioner} sound over five severity levels using Wav2vec2 Base with greedy decoding. SNRs (dB) are listed in the first row. The best results are bold.}
\label{subtable:aircon_main}
\end{minipage}%
\hfill
\begin{minipage}{0.47\textwidth}
\centering
    \begin{tabular}{cccccc}
    \toprule
     & 10    & 5     & 0     & -5    & -10 \\
    \midrule
    Source & 26.2  & 34.0  & 44.4  & 56.4  & 69.0 \\
    Tent  & 21.0  & 27.9  & 37.0  & 49.2  & 63.0 \\
    SAR   & 23.0  & 30.3  & 39.7  & 52.1  & 65.3 \\
    TeCo  & 21.0  & 27.8  & 37.0  & 49.1  & 63.0 \\
    SUTA  & 17.9  & 23.3  & 30.4  & 41.0  & 53.4 \\
    Ours  & \textbf{17.5}  & \textbf{22.8}  & \textbf{29.9}  & \textbf{40.4}  & \textbf{52.6} \\
    \bottomrule
    \end{tabular}%
\caption{WER (\%) results on \textbf{Typing} sound over five severity levels using Wav2vec2 Base with greedy decoding. SNRs (dB) are listed in the first row. The best results are bold.}
\label{subtable:typing_main}
\end{minipage}

\end{table}

\begin{table*}[htbp]
\begin{center}
\begin{tabular}{cccccc|cccccccccccccccccccccccc}
\toprule[1pt]
\multicolumn{6}{c}{\multirow{2}{*}{Method}} & \multicolumn{6}{c}{\multirow{1}{*}{DSing-dev}} & \multicolumn{6}{c}{\multirow{1}{*}{DSing-test}} & \multicolumn{6}{c}{\multirow{1}{*}{Hansen}}  & \multicolumn{6}{c}{\multirow{1}{*}{Average}} \\  \cline{7-30}
\multicolumn{6}{c}{} & \multicolumn{3}{c}{\multirow{1}{*}{Base}} & \multicolumn{3}{c}{\multirow{1}{*}{Large}} & \multicolumn{3}{c}{\multirow{1}{*}{Base}} & \multicolumn{3}{c}{\multirow{1}{*}{Large}} & \multicolumn{3}{c}{\multirow{1}{*}{Base}} & \multicolumn{3}{c}{\multirow{1}{*}{Large}}  &  \multicolumn{3}{c}{\multirow{1}{*}{Base}} & \multicolumn{3}{c}{\multirow{1}{*}{Large}} \\ 
\midrule[0.5pt]
 \multicolumn{6}{c}{Greedy Search} & \multicolumn{6}{c}{} & \multicolumn{6}{c}{} & \multicolumn{6}{c}{} & \multicolumn{6}{c}{} \\
 
\midrule[0.5pt]
 \multicolumn{6}{c}{Source} & \multicolumn{3}{c}{\multirow{1}{*}{$61.8$}} & \multicolumn{3}{c}{\multirow{1}{*}{$40.6$}} & \multicolumn{3}{c}{\multirow{1}{*}{$60.1$}} & \multicolumn{3}{c}{\multirow{1}{*}{$38.8$}} & \multicolumn{3}{c}{\multirow{1}{*}{$64.3$}} & \multicolumn{3}{c}{\multirow{1}{*}{$43.7$}} & \multicolumn{3}{c}{\multirow{1}{*}{$62.1$}} & \multicolumn{3}{c}{\multirow{1}{*}{$41.0$}}  \\ 
 
 \multicolumn{6}{c}{Tent} & \multicolumn{3}{c}{\multirow{1}{*}{$55.7$}} & \multicolumn{3}{c}{\multirow{1}{*}{$34.8$}} & \multicolumn{3}{c}{\multirow{1}{*}{$56.1$}} & \multicolumn{3}{c}{\multirow{1}{*}{$33.2$}} & \multicolumn{3}{c}{\multirow{1}{*}{$60.2$}} & \multicolumn{3}{c}{\multirow{1}{*}{$39.1$}} & \multicolumn{3}{c}{\multirow{1}{*}{$57.3$}} & \multicolumn{3}{c}{\multirow{1}{*}{$35.7$}}  \\ 
 
 \multicolumn{6}{c}{SAR} & \multicolumn{3}{c}{\multirow{1}{*}{$58.8$}} & \multicolumn{3}{c}{\multirow{1}{*}{$40.6$}} & \multicolumn{3}{c}{\multirow{1}{*}{$57.2$}} & \multicolumn{3}{c}{\multirow{1}{*}{$38.2$}} & \multicolumn{3}{c}{\multirow{1}{*}{$62.7$}} & \multicolumn{3}{c}{\multirow{1}{*}{$42.7$}} & \multicolumn{3}{c}{\multirow{1}{*}{$59.6$}} & \multicolumn{3}{c}{\multirow{1}{*}{$40.5$}}  \\ 
 
 \multicolumn{6}{c}{TeCo} & \multicolumn{3}{c}{\multirow{1}{*}{$56.2$}} & \multicolumn{3}{c}{\multirow{1}{*}{$35.0$}} & \multicolumn{3}{c}{\multirow{1}{*}{$55.6$}} & \multicolumn{3}{c}{\multirow{1}{*}{$33.1$}} & \multicolumn{3}{c}{\multirow{1}{*}{$60.0$}} & \multicolumn{3}{c}{\multirow{1}{*}{$39.1$}} & \multicolumn{3}{c}{\multirow{1}{*}{$57.3$}} & \multicolumn{3}{c}{\multirow{1}{*}{$35.7$}}  \\ 
 
 \multicolumn{6}{c}{SUTA} & \multicolumn{3}{c}{\multirow{1}{*}{$53.9$}} & \multicolumn{3}{c}{\multirow{1}{*}{$34.9$}} & \multicolumn{3}{c}{\multirow{1}{*}{$51.3$}} & \multicolumn{3}{c}{\multirow{1}{*}{$33.6$}} & \multicolumn{3}{c}{\multirow{1}{*}{$\textbf{58.0}$}} & \multicolumn{3}{c}{\multirow{1}{*}{$39.3$}} & \multicolumn{3}{c}{\multirow{1}{*}{$54.4$}} & \multicolumn{3}{c}{\multirow{1}{*}{$35.9$}}  \\ 
 \multicolumn{6}{c}{Ours} & \multicolumn{3}{c}{\multirow{1}{*}{$\textbf{53.5}$}} & \multicolumn{3}{c}{\multirow{1}{*}{$\textbf{34.0}$}} & \multicolumn{3}{c}{\multirow{1}{*}{$\textbf{50.1}$}} & \multicolumn{3}{c}{\multirow{1}{*}{$\textbf{31.2}$}} & \multicolumn{3}{c}{\multirow{1}{*}{$\textbf{58.0}$}} & \multicolumn{3}{c}{\multirow{1}{*}{$\textbf{37.9}$}} & \multicolumn{3}{c}{\multirow{1}{*}{$\textbf{53.9}$}} & \multicolumn{3}{c}{\multirow{1}{*}{$\textbf{34.4}$}}  \\ 

\midrule[0.5pt]
\multicolumn{6}{c}{Beam Search} & \multicolumn{6}{c}{} & \multicolumn{6}{c}{} & \multicolumn{6}{c}{} & \multicolumn{6}{c}{} \\
\midrule[0.5pt]
 
 \multicolumn{6}{c}{Source+LM} & \multicolumn{3}{c}{\multirow{1}{*}{$58.6$}} & \multicolumn{3}{c}{\multirow{1}{*}{$41.1$}} & \multicolumn{3}{c}{\multirow{1}{*}{$55.3$}} & \multicolumn{3}{c}{\multirow{1}{*}{$37.6$}} & \multicolumn{3}{c}{\multirow{1}{*}{$60.1$}} & \multicolumn{3}{c}{\multirow{1}{*}{$43.5$}} & \multicolumn{3}{c}{\multirow{1}{*}{$58.0$}} & \multicolumn{3}{c}{\multirow{1}{*}{$40.7$}}  \\ 
 
 \multicolumn{6}{c}{SGEM} & \multicolumn{3}{c}{\multirow{1}{*}{$54.4$}} & \multicolumn{3}{c}{\multirow{1}{*}{$34.4$}} & \multicolumn{3}{c}{\multirow{1}{*}{$50.8$}} & \multicolumn{3}{c}{\multirow{1}{*}{$33.0$}} & \multicolumn{3}{c}{\multirow{1}{*}{$57.8$}} & \multicolumn{3}{c}{\multirow{1}{*}{$38.6$}} & \multicolumn{3}{c}{\multirow{1}{*}{$54.3$}} & \multicolumn{3}{c}{\multirow{1}{*}{$35.3$}}  \\ 

 \multicolumn{6}{c}{Ours+LM} & \multicolumn{3}{c}{\multirow{1}{*}{$\textbf{53.2}$}} & \multicolumn{3}{c}{\multirow{1}{*}{$\textbf{33.3}$}} & \multicolumn{3}{c}{\multirow{1}{*}{$\textbf{50.0}$}} & \multicolumn{3}{c}{\multirow{1}{*}{$\textbf{30.3}$}} & \multicolumn{3}{c}{\multirow{1}{*}{$\textbf{57.7}$}} & \multicolumn{3}{c}{\multirow{1}{*}{$\textbf{37.5}$}} & \multicolumn{3}{c}{\multirow{1}{*}{$\textbf{53.6}$}} & \multicolumn{3}{c}{\multirow{1}{*}{$\textbf{33.7}$}}  \\ 

\bottomrule[1pt]
\end{tabular}
\caption{WER (\%) results on DSing-dev, DSing-test, and Hansen with greedy search and beam search. Base and Large denote Wav2vec2 Base and Wav2vec2 Large respectively. The best results are bold. }
\label{table:sing}
\vspace{-3mm}
\end{center}
\end{table*}

\subsection{Robustness to Synthetic Noises}
\label{noise}

\paragraph{Gaussian Noises. } In the initial phase of our experiments, we focus on synthetic data and assess the robustness in the presence of various levels of Gaussian noise injected into the test speech audio. The outcomes are reported in Table~\ref{table:ls}. It is observed that our proposed method consistently outperforms existing baseline approaches across five levels of noise. Notably, our approach achieves a relative improvement of $21.5\%$ on average in terms of WER, when compared to using the source model without adaptation. 

Furthermore, it is imperative to note that SAR, designed for addressing wild vision data shifts, demonstrates comparatively less improvement compared with the Tent method. This observation underscores the limitations of filtering noisy frames for speech recognition. Instead, the learning-based adaptation adopted in our method shows superiority. Moreover, we discover that TeCo provides marginal improvement compared to Tent, indicating that coherence regularization is limited in the context of noisy frames. In contrast, our confidence enhanced adaptation yields further benefits for temporal consistency regularization.

\paragraph{Environmental Sounds. } We further evaluate the robustness on LS-P, which introduces eight common environmental sounds in the test audio at five levels of severity. The results of adding Air Conditioner sound and Typing sound are reported in Table \ref{subtable:aircon_main} and Table \ref{subtable:typing_main} respectively (Full experimental results can be found in Appendix~\ref{app:ls-p}). It is noticeable that our method can yield over $30\%$ relative improvements in low-SNR scenarios. Notably, for the case with 5 dB SNR in Table \ref{subtable:aircon_main}, our method demonstrates a substantial $41.7\%$ relative improvement, suggesting its efficacy in mitigating the impact of real-world environmental sound corruption.

\subsection{Robustness to Real-World Data Shifts}
\label{realworld}
\paragraph{L2 Accents. } Data shifts resulting from accent variations are a common occurrence in real-world scenarios, arising from differences in dialects or non-native speech patterns. Another pertinent instance of such shifts is encountered in children's speech, which is also a common pronunciation change and one type of accent in the real world. In order to assess the robustness to such pronunciation variations, we undertake the test-time adaptation to accents exhibited by L2 learners using the L2-Arctic dataset. To comprehensively evaluate the performance, we evaluate all speakers for each L1 and present the speaker-level results for each L1 in Appendix~\ref{app:l2}. The experimental findings consistently underscore the superiority of our proposed method across different L1 categories.

\paragraph{Singing Voice.} In this session, we discuss the robustness of ASR fine-tuned acoustic foundation models to singing voice for the first time. Singing, also referred to as sung speech, is characterized by a distinctive pronunciation pattern. Notably, it encompasses various frequency fluctuations, including the apparent pitch variations along with the melody. This constitutes a tremendous covariate shift, rendering the adaptation from speech to singing more challenging than that from speech to speech. Moreover, the existence of professional singing techniques further compounds the challenges associated with adaptation. For instance, the elongation of word pronunciation, a common occurrence in singing, is a departure from typical speech patterns. 

To evaluate the adaptation performance under shifts from singing voice, we conduct experiments on three datasets, utilizing both Wav2vec2 Base and Wav2vec2 Large models. The outcomes are presented in Table~\ref{table:sing}. The results indicate that our proposed method consistently attains the best performances for both Base and Large models. In addition, the Wav2vec2 Large model exhibits superior robustness than the Base model. Nevertheless, it still experiences a noticeable performance degradation when compared with adaptation in noise and accent robustness evaluations, suggesting the limited ability of acoustic foundation models under wild acoustic test settings. 

\begin{table}[t]

\centering
\begin{tabular}{cccccccccccccccccc}
\toprule[1pt]
 \multicolumn{6}{c}{Method} & \multicolumn{6}{c}{Conformer} & \multicolumn{6}{c}{Transducer} \\ 
 \midrule[0.1pt]
 \multicolumn{6}{c}{Source} & \multicolumn{6}{c}{$62.2$} & \multicolumn{6}{c}{$48.8$}  \\ 
\multicolumn{6}{c}{SUTA} & \multicolumn{6}{c}{$55.9$} & \multicolumn{6}{c}{$44.8$}   \\ 
\multicolumn{6}{c}{SGEM} & \multicolumn{6}{c}{$55.7$} & \multicolumn{6}{c}{$44.5$}   \\ 
\multicolumn{6}{c}{Ours} & \multicolumn{6}{c}{$\textbf{55.4}$} & \multicolumn{6}{c}{$\textbf{43.0}$}   \\ 
\bottomrule[1pt]
\end{tabular}
\caption{WER (\%) results on DSing-test using Conformer-CTC and Conformer-Transducer.}
\label{subtable:abl_diff_asr}
\vspace{-3mm}

\end{table}

\subsection{Decoding Strategies}
\label{decoding}

We discuss the decoding strategies employed in experiments in this session. In our preceding experiments, we mainly utilize greedy decoding, which does not explicitly tackle the text-domain changes. In the subsequent analysis, we compare our proposed method with SGEM, which leverages beam search for decoding. The results are presented in Table~\ref{table:sing}. Notably, our findings reveal that even in the absence of explicit adaptation for the language model, our approach still consistently outperforms SGEM. We also observe that the results achieved by our method using greedy search can, on average, surpass those of SGEM. We conjecture that our proposed short-term consistency regularization addresses the label shift implicitly by fostering label coherency among neighbor frames. Moreover, it is discovered that the enhancements facilitated by adaptation are more pronounced compared to the ones achieved through beam search, indicating the significance of test-time adaptation for acoustic foundation models.

\section{Analysis}

\subsection{Generalization on Different ASR Models}

\label{app:diff_asr_models}
We examine the robustness of CTC-based acoustic foundation models in our main experiments and Appendix~\ref{app: more_models}. To verify the efficacy of our method on other end-to-end ASR models such as Conformer and Transducer, we conducted experiments on Conformer-CTC~\citep{conformer} and Conformer-Transducer~\cite{conf_trans} as per~\citet{sgem}. For consistent setting and fair comparison, we experimented with DSing-test and reported the results in Table~\ref{subtable:abl_diff_asr}. The empirical results illustrate that our proposed method can be generalized to different end-to-end ASR models and outperform SUTA and SGEM baselines.

\subsection{Ablation Study}
We conduct the ablation study on Noise, Accent, Singing shifts respectively using Wav2vec2 Base with greedy search to dissect the individual impact of two core components proposed in our methods. The results presented in Table \ref{subtable:abl_comp} illustrate that the removal of short-term consistency regularization (STCR) leads to a relatively modest decline in performance, in contrast to the more substantial deterioration observed upon the removal of confidence enhanced adaptation (CEA). This observation underscores the significance of our proposed CEA. Furthermore, the introduction of STCR yields additional performance gains when employed in conjunction with CEA. These experimental findings also indicate a pronounced efficacy of our method in mitigating noise shifts as opposed to accent and singing shifts. We conjecture the reason could be that the shift caused by Gaussian noises for each frame is consistent while other shifts such as accent shift could be different within frames. 

 


\begin{table}

\centering
\begin{tabular}{ccccccccccccccc}
\toprule[1pt]
 \multicolumn{6}{c}{Method} & \multicolumn{3}{c}{Noise} & \multicolumn{3}{c}{Accent} & \multicolumn{3}{c}{Singing} \\ 
 \midrule[0.5pt]
 \multicolumn{6}{l}{Ours} & \multicolumn{3}{c}{$\textbf{24.0}$} & \multicolumn{3}{c}{$\textbf{23.0}$}  & \multicolumn{3}{c}{$\textbf{50.1}$}  \\ 
\multicolumn{6}{l}{w/o STCR} & \multicolumn{3}{c}{$25.1$} & \multicolumn{3}{c}{$23.4$}  & \multicolumn{3}{c}{$51.0$}   \\ 
\multicolumn{6}{l}{w/o CEA} & \multicolumn{3}{c}{$35.9$} & \multicolumn{3}{c}{$26.9$}  & \multicolumn{3}{c}{$54.5$}   \\ 
\bottomrule[1pt]
\end{tabular}
\caption{Ablation study of core components proposed in our work. WER (\%) results are reported.}
\label{subtable:abl_comp}

\vspace{-3mm}
\end{table}
\begin{table*}[t]
\centering
\begin{adjustbox}{max width=\textwidth}
\begin{tabular}{lccccc}
\toprule
\textbf{Model} & \textbf{Level 1} & \textbf{Level 2} & \textbf{Level 3} & \textbf{Level 4} & \textbf{Level 5} \\
\midrule
Whisper-Base & 20.7 & 25.6 & 30.1 & 36.6 & 50.3 \\
Whisper-Base.en & 13.9 & 20.1 & 22.2 & 26.6 & 36.8 \\
\bottomrule
\end{tabular}
\end{adjustbox}

\caption{WER (\%) results on LS-C over five severity levels using Whisper-Base and Whisper-Base.en.}
\label{whisper_comp}
\end{table*}
\subsection{Latency Analysis}
We did the adaptation with a single coming utterance and counted the difference between the time when the utterance has ended and the time when the adaptation process has ended. We calculate the average latency over all samples of Librispeech test-other set on Wav2vec2 Base and obtain the latency of $1.07$ seconds. The average recognition run-time on A5000 is $1.20$ seconds. We believe this could be an acceptable delay due to large parameter sizes for acoustic foundation models. We provide additional comparisons in terms of computing in Appendix~\ref{compute_time}.

\subsection{Comparison with Whisper}
State-of-the-art models such as Whisper~\citep{whisper} improve noise robustness by leveraging a large training corpus with data augmentation by adding noise. To gain an insight on how our TTA method for improving noise robustness compares with Whisper, we conduct additional experiments on LS-C using Whisper and report the performance in Table~\ref{whisper_comp}. We choose Whisper-Base due to its comparable parameter size to Wav2vec2 Base. Note that it is impossible to make a fair comparison since both Whisper Base (74M) and Small (244M) have different parameter sizes to the Wav2vec2 Base (~90M). It is interesting to observe that the performances of adapted Wav2vec2 Base in Table~\ref{table:ls} can surpass those of unadapted Whisper-Base for severity levels 1 to 4, and the unadapted Whisper-Base.en for severity levels 1 and 2, demonstrating the effectiveness of the proposed TTA method. This also indicates that training with augmented data like Whisper brings more robustness to more severe corruption. However, whether these results generalize to wilder acoustic test settings, which are beyond the scope of Whisper's objective but central to ours, remains an open question for future investigation.



\section{Conclusions}
In this paper, we study the Test-Time Adaptation of ASR fine-tuned acoustic foundation models under wild acoustic test settings. By investigating the role of high-entropy noisy frames within non-silent speech segments, we introduce Confidence Enhanced Adaptation with a confidence-aware weight optimization scheme to prioritize these noisy frames for efficient adaptation via denoising their intermediate representations rather than discarding them. Moreover, our emphasis on short-term stability of speech signals leads us to apply consistency regularization, yielding further improvement for stable online TTA. Our experimental findings suggest a consistent improvement for different types of acoustic shifts and different degrees of corruption on synthetic and real-world datasets, demonstrating the efficacy of our approach under wild acoustic test settings.


\section*{Limitations}

Our work is subject to several limitations. Firstly, further research endeavors could encompass a broader exploration of adaptation techniques for the decoder model, particularly for text-domain adaptation. It remains challenging to adapt language models to address text-domain shifts due to the unavailability of target domain texts in the TTA setting. Additionally, we mainly experiment with ASR fine-tuned acoustic foundation models. The broader applicability of our method to diverse speech tasks, including but not limited to multi-speaker-related scenarios, spoken language understanding, and general audio classification tasks remains unexplored. Therefore, we consider adapting our approach to these tasks under wild acoustic test settings as the future work. Finally, our work is not specifically tailored for online streaming applications and TTA under streaming scenarios for latency reduction is definitely essential in future work.  



\section*{Acknowledgements}
The authors would like to thank anonymous reviewers for their valuable suggestions. This project is funded by a research grant MOE-MOESOL2021-0005 from the Ministry of Education in Singapore.


\newpage
\bibliography{custom}

\begin{thebibliography}{48}
\expandafter\ifx\csname natexlab\endcsname\relax\def\natexlab#1{#1}\fi

\bibitem[{Baevski et~al.(2020)Baevski, Zhou, Mohamed, and Auli}]{wav2vec2}
Alexei Baevski, Yuhao Zhou, Abdelrahman Mohamed, and Michael Auli. 2020.
\newblock wav2vec 2.0: A framework for self-supervised learning of speech representations.
\newblock \emph{Advances in neural information processing systems}, 33:12449--12460.

\bibitem[{Bell et~al.(2020)Bell, Fainberg, Klejch, Li, Renals, and Swietojanski}]{adaptation_overview}
Peter Bell, Joachim Fainberg, Ondrej Klejch, Jinyu Li, Steve Renals, and Pawel Swietojanski. 2020.
\newblock Adaptation algorithms for neural network-based speech recognition: An overview.
\newblock \emph{IEEE Open Journal of Signal Processing}, 2:33--66.

\bibitem[{Burchi and Vielzeuf(2021)}]{conf_trans}
Maxime Burchi and Valentin Vielzeuf. 2021.
\newblock Efficient conformer: Progressive downsampling and grouped attention for automatic speech recognition.
\newblock In \emph{2021 IEEE Automatic Speech Recognition and Understanding Workshop (ASRU)}, pages 8--15. IEEE.

\bibitem[{Chen et~al.(2022)Chen, Wang, Chen, Wu, Liu, Chen, Li, Kanda, Yoshioka, Xiao et~al.}]{wavlm}
Sanyuan Chen, Chengyi Wang, Zhengyang Chen, Yu~Wu, Shujie Liu, Zhuo Chen, Jinyu Li, Naoyuki Kanda, Takuya Yoshioka, Xiong Xiao, et~al. 2022.
\newblock Wavlm: Large-scale self-supervised pre-training for full stack speech processing.
\newblock \emph{IEEE Journal of Selected Topics in Signal Processing}, 16(6):1505--1518.

\bibitem[{Dabike and Barker(2019)}]{dsing}
Gerardo~Roa Dabike and Jon Barker. 2019.
\newblock Automatic lyric transcription from karaoke vocal tracks: Resources and a baseline system.
\newblock In \emph{Interspeech}, pages 579--583.

\bibitem[{Deng et~al.(2022)Deng, Xie, Wang, Cui, Xue, Jin, Geng, Li, Liu, and Meng}]{speaker2}
Jiajun Deng, Xurong Xie, Tianzi Wang, Mingyu Cui, Boyang Xue, Zengrui Jin, Mengzhe Geng, Guinan Li, Xunying Liu, and Helen Meng. 2022.
\newblock Confidence score based conformer speaker adaptation for speech recognition.
\newblock \emph{arXiv preprint arXiv:2206.12045}.

\bibitem[{Gong et~al.(2022)Gong, Jeong, Kim, Kim, Shin, and Lee}]{note}
Taesik Gong, Jongheon Jeong, Taewon Kim, Yewon Kim, Jinwoo Shin, and Sung-Ju Lee. 2022.
\newblock Note: Robust continual test-time adaptation against temporal correlation.
\newblock \emph{Advances in Neural Information Processing Systems}, 35:27253--27266.

\bibitem[{Gulati et~al.(2020)Gulati, Qin, Chiu, Parmar, Zhang, Yu, Han, Wang, Zhang, Wu et~al.}]{conformer}
Anmol Gulati, James Qin, Chung-Cheng Chiu, Niki Parmar, Yu~Zhang, Jiahui Yu, Wei Han, Shibo Wang, Zhengdong Zhang, Yonghui Wu, et~al. 2020.
\newblock Conformer: Convolution-augmented transformer for speech recognition.
\newblock \emph{arXiv preprint arXiv:2005.08100}.

\bibitem[{Hansen and Fraunhofer(2012)}]{hansen}
Jens~Kofod Hansen and IDMT Fraunhofer. 2012.
\newblock Recognition of phonemes in a-cappella recordings using temporal patterns and mel frequency cepstral coefficients.
\newblock In \emph{9th Sound and Music Computing Conference (SMC)}, pages 494--499.

\bibitem[{Hendrycks and Dietterich(2019)}]{benchmarking}
Dan Hendrycks and Thomas Dietterich. 2019.
\newblock Benchmarking neural network robustness to common corruptions and perturbations.
\newblock \emph{arXiv preprint arXiv:1903.12261}.

\bibitem[{Hou et~al.(2021)Hou, Wang, Tan, Qin, and Shinozaki}]{feature_align}
Wenxin Hou, Jindong Wang, Xu~Tan, Tao Qin, and Takahiro Shinozaki. 2021.
\newblock Cross-domain speech recognition with unsupervised character-level distribution matching.
\newblock \emph{arXiv preprint arXiv:2104.07491}.

\bibitem[{Hsu et~al.(2021)Hsu, Bolte, Tsai, Lakhotia, Salakhutdinov, and Mohamed}]{hubert}
Wei-Ning Hsu, Benjamin Bolte, Yao-Hung~Hubert Tsai, Kushal Lakhotia, Ruslan Salakhutdinov, and Abdelrahman Mohamed. 2021.
\newblock Hubert: Self-supervised speech representation learning by masked prediction of hidden units.
\newblock \emph{IEEE/ACM Transactions on Audio, Speech, and Language Processing}, 29:3451--3460.

\bibitem[{Hsu et~al.(2017)Hsu, Zhang, and Glass}]{dataaug}
Wei-Ning Hsu, Yu~Zhang, and James Glass. 2017.
\newblock Unsupervised domain adaptation for robust speech recognition via variational autoencoder-based data augmentation.
\newblock In \emph{2017 IEEE automatic speech recognition and understanding workshop (ASRU)}, pages 16--23. IEEE.

\bibitem[{Huang et~al.(2022)Huang, Gu, Wang, Xiao, Liu, and Wang}]{huang2022extrapolative}
Hengguan Huang, Xiangming Gu, Hao Wang, Chang Xiao, Hongfu Liu, and Ye~Wang. 2022.
\newblock Extrapolative continuous-time bayesian neural network for fast training-free test-time adaptation.
\newblock \emph{Advances in Neural Information Processing Systems}, 35:36000--36013.

\bibitem[{Huang et~al.(2021)Huang, Liu, Wang, Xiao, and Wang}]{huang2021strode}
Hengguan Huang, Hongfu Liu, Hao Wang, Chang Xiao, and Ye~Wang. 2021.
\newblock Strode: Stochastic boundary ordinary differential equation.
\newblock In \emph{International Conference on Machine Learning}, pages 4435--4445. PMLR.

\bibitem[{Huang and Mak(2017)}]{huang2017improve}
Hengguan Huang and Brian Mak. 2017.
\newblock To improve the robustness of lstm-rnn acoustic models using higher-order feedback from multiple histories.
\newblock In \emph{INTERSPEECH}, pages 3862--3866.

\bibitem[{Huang and Mak(2018)}]{huang2018wavenet}
Hengguan Huang and Brian Mak. 2018.
\newblock Wavenet mh-sru: Deep and wide multiple-history simple recurrent unit for speech recognition.
\newblock In \emph{2018 11th International Symposium on Chinese Spoken Language Processing (ISCSLP)}, pages 141--145. IEEE.

\bibitem[{Huang et~al.(2019)Huang, Wang, and Mak}]{huang2019recurrent}
Hengguan Huang, Hao Wang, and Brian Mak. 2019.
\newblock Recurrent poisson process unit for speech recognition.
\newblock In \emph{Proceedings of the AAAI Conference on Artificial Intelligence}, volume~33, pages 6538--6545.

\bibitem[{Ioffe and Szegedy(2015)}]{bn}
Sergey Ioffe and Christian Szegedy. 2015.
\newblock Batch normalization: Accelerating deep network training by reducing internal covariate shift.
\newblock In \emph{International conference on machine learning}, pages 448--456. pmlr.

\bibitem[{Kim et~al.(2023)Kim, Park, Shim, and Yang}]{sgem}
Changhun Kim, Joonhyung Park, Hajin Shim, and Eunho Yang. 2023.
\newblock Sgem: Test-time adaptation for automatic speech recognition via sequential-level generalized entropy minimization.
\newblock \emph{arXiv preprint arXiv:2306.01981}.

\bibitem[{Koh et~al.(2021)Koh, Sagawa, Marklund, Xie, Zhang, Balsubramani, Hu, Yasunaga, Phillips, Gao et~al.}]{wilds}
Pang~Wei Koh, Shiori Sagawa, Henrik Marklund, Sang~Michael Xie, Marvin Zhang, Akshay Balsubramani, Weihua Hu, Michihiro Yasunaga, Richard~Lanas Phillips, Irena Gao, et~al. 2021.
\newblock Wilds: A benchmark of in-the-wild distribution shifts.
\newblock In \emph{International Conference on Machine Learning}, pages 5637--5664. PMLR.

\bibitem[{K{\"u}rzinger et~al.(2020)K{\"u}rzinger, Winkelbauer, Li, Watzel, and Rigoll}]{kurzinger2020ctc}
Ludwig K{\"u}rzinger, Dominik Winkelbauer, Lujun Li, Tobias Watzel, and Gerhard Rigoll. 2020.
\newblock Ctc-segmentation of large corpora for {German} end-to-end speech recognition.
\newblock In \emph{International Conference on Speech and Computer}, pages 267--278. Springer.

\bibitem[{Li et~al.(2014)Li, Deng, Gong, and Haeb-Umbach}]{speechrob2}
Jinyu Li, Li~Deng, Yifan Gong, and Reinhold Haeb-Umbach. 2014.
\newblock An overview of noise-robust automatic speech recognition.
\newblock \emph{IEEE/ACM Transactions on Audio, Speech, and Language Processing}, 22(4):745--777.

\bibitem[{Li et~al.(2017)Li, Seltzer, Wang, Zhao, and Gong}]{distill}
Jinyu Li, Michael~L Seltzer, Xi~Wang, Rui Zhao, and Yifan Gong. 2017.
\newblock Large-scale domain adaptation via teacher-student learning.
\newblock \emph{arXiv preprint arXiv:1708.05466}.

\bibitem[{Lim et~al.(2023)Lim, Kim, Choo, and Choi}]{ttn}
Hyesu Lim, Byeonggeun Kim, Jaegul Choo, and Sungha Choi. 2023.
\newblock Ttn: A domain-shift aware batch normalization in test-time adaptation.
\newblock \emph{arXiv preprint arXiv:2302.05155}.

\bibitem[{Lin et~al.(2022)Lin, Li, and Lee}]{suta}
Guan-Ting Lin, Shang-Wen Li, and Hung-yi Lee. 2022.
\newblock Listen, adapt, better wer: Source-free single-utterance test-time adaptation for automatic speech recognition.
\newblock \emph{arXiv preprint arXiv:2203.14222}.

\bibitem[{Niu et~al.(2022)Niu, Wu, Zhang, Chen, Zheng, Zhao, and Tan}]{eata}
Shuaicheng Niu, Jiaxiang Wu, Yifan Zhang, Yaofo Chen, Shijian Zheng, Peilin Zhao, and Mingkui Tan. 2022.
\newblock Efficient test-time model adaptation without forgetting.
\newblock In \emph{International conference on machine learning}, pages 16888--16905. PMLR.

\bibitem[{Niu et~al.(2023)Niu, Wu, Zhang, Wen, Chen, Zhao, and Tan}]{sar}
Shuaicheng Niu, Jiaxiang Wu, Yifan Zhang, Zhiquan Wen, Yaofo Chen, Peilin Zhao, and Mingkui Tan. 2023.
\newblock Towards stable test-time adaptation in dynamic wild world.
\newblock \emph{arXiv preprint arXiv:2302.12400}.

\bibitem[{Panayotov et~al.(2015)Panayotov, Chen, Povey, and Khudanpur}]{librispeech}
Vassil Panayotov, Guoguo Chen, Daniel Povey, and Sanjeev Khudanpur. 2015.
\newblock Librispeech: an asr corpus based on public domain audio books.
\newblock In \emph{2015 IEEE international conference on acoustics, speech and signal processing (ICASSP)}, pages 5206--5210. IEEE.

\bibitem[{Rabiner et~al.(2007)Rabiner, Schafer et~al.}]{rabiner2007introduction}
Lawrence~R Rabiner, Ronald~W Schafer, et~al. 2007.
\newblock Introduction to digital speech processing.
\newblock \emph{Foundations and Trends{\textregistered} in Signal Processing}, 1(1--2):1--194.

\bibitem[{Radford et~al.(2023)Radford, Kim, Xu, Brockman, McLeavey, and Sutskever}]{whisper}
Alec Radford, Jong~Wook Kim, Tao Xu, Greg Brockman, Christine McLeavey, and Ilya Sutskever. 2023.
\newblock Robust speech recognition via large-scale weak supervision.
\newblock In \emph{International Conference on Machine Learning}, pages 28492--28518. PMLR.

\bibitem[{Reddy et~al.(2019)Reddy, Beyrami, Pool, Cutler, Srinivasan, and Gehrke}]{noisyenv}
Chandan~KA Reddy, Ebrahim Beyrami, Jamie Pool, Ross Cutler, Sriram Srinivasan, and Johannes Gehrke. 2019.
\newblock A scalable noisy speech dataset and online subjective test framework.
\newblock \emph{Proc. Interspeech 2019}, pages 1816--1820.

\bibitem[{Sun et~al.(2018)Sun, Yeh, Hwang, Ostendorf, and Xie}]{adver2}
Sining Sun, Ching-Feng Yeh, Mei-Yuh Hwang, Mari Ostendorf, and Lei Xie. 2018.
\newblock Domain adversarial training for accented speech recognition.
\newblock In \emph{2018 IEEE international conference on acoustics, speech and signal processing (ICASSP)}, pages 4854--4858. IEEE.

\bibitem[{Sun et~al.(2017)Sun, Zhang, Xie, and Zhang}]{adver1}
Sining Sun, Binbin Zhang, Lei Xie, and Yanning Zhang. 2017.
\newblock An unsupervised deep domain adaptation approach for robust speech recognition.
\newblock \emph{Neurocomputing}, 257:79--87.

\bibitem[{Swietojanski et~al.(2016)Swietojanski, Li, and Renals}]{unsupadapt}
Pawel Swietojanski, Jinyu Li, and Steve Renals. 2016.
\newblock Learning hidden unit contributions for unsupervised acoustic model adaptation.
\newblock \emph{IEEE/ACM Transactions on Audio, Speech, and Language Processing}, 24(8):1450--1463.

\bibitem[{Swietojanski and Renals(2014)}]{speakadapt}
Pawel Swietojanski and Steve Renals. 2014.
\newblock Learning hidden unit contributions for unsupervised speaker adaptation of neural network acoustic models.
\newblock In \emph{2014 IEEE Spoken Language Technology Workshop (SLT)}, pages 171--176. IEEE.

\bibitem[{Wang et~al.(2020)Wang, Shelhamer, Liu, Olshausen, and Darrell}]{tent}
Dequan Wang, Evan Shelhamer, Shaoteng Liu, Bruno Olshausen, and Trevor Darrell. 2020.
\newblock Tent: Fully test-time adaptation by entropy minimization.
\newblock \emph{arXiv preprint arXiv:2006.10726}.

\bibitem[{Wang et~al.(2022)Wang, Fink, Van~Gool, and Dai}]{cota}
Qin Wang, Olga Fink, Luc Van~Gool, and Dengxin Dai. 2022.
\newblock Continual test-time domain adaptation.
\newblock In \emph{Proceedings of the IEEE/CVF Conference on Computer Vision and Pattern Recognition}, pages 7201--7211.

\bibitem[{Wei et~al.(2022)Wei, Huang, Gu, Wang, and Wang}]{wei2022unsupervised}
Wei Wei, Hengguan Huang, Xiangming Gu, Hao Wang, and Ye~Wang. 2022.
\newblock Unsupervised mismatch localization in cross-modal sequential data with application to mispronunciations localization.
\newblock \emph{Transactions on Machine Learning Research}.

\bibitem[{Yang et~al.(2023{\natexlab{a}})Yang, Li, Zhang, Chen, Prabhavalkar, Sainath, and Strohman}]{reproxl}
Chao-Han~Huck Yang, Bo~Li, Yu~Zhang, Nanxin Chen, Rohit Prabhavalkar, Tara~N Sainath, and Trevor Strohman. 2023{\natexlab{a}}.
\newblock From english to more languages: Parameter-efficient model reprogramming for cross-lingual speech recognition.
\newblock In \emph{ICASSP 2023-2023 IEEE International Conference on Acoustics, Speech and Signal Processing (ICASSP)}, pages 1--5. IEEE.

\bibitem[{Yang et~al.(2021)Yang, Tsai, and Chen}]{reprogramming}
Chao-Han~Huck Yang, Yun-Yun Tsai, and Pin-Yu Chen. 2021.
\newblock Voice2series: Reprogramming acoustic models for time series classification.
\newblock In \emph{International conference on machine learning}, pages 11808--11819. PMLR.

\bibitem[{Yang et~al.(2023{\natexlab{b}})Yang, Yang, and Chien}]{reproadapter}
Li-Jen Yang, Chao-Han~Huck Yang, and Jen-Tzung Chien. 2023{\natexlab{b}}.
\newblock Parameter-efficient learning for text-to-speech accent adaptation.
\newblock \emph{arXiv preprint arXiv:2305.11320}.

\bibitem[{Yang et~al.(2023{\natexlab{c}})Yang, Yang, Guo, Yao, Kang, Kuang, Lin, Chen, and Povey}]{yang2023blank}
Yifan Yang, Xiaoyu Yang, Liyong Guo, Zengwei Yao, Wei Kang, Fangjun Kuang, Long Lin, Xie Chen, and Daniel Povey. 2023{\natexlab{c}}.
\newblock Blank-regularized ctc for frame skipping in neural transducer.
\newblock \emph{arXiv preprint arXiv:2305.11558}.

\bibitem[{Yi et~al.(2023)Yi, Yang, Wang, Li, Tan, and Kot}]{teco}
Chenyu Yi, Siyuan Yang, Yufei Wang, Haoliang Li, Yap-Peng Tan, and Alex~C Kot. 2023.
\newblock Temporal coherent test-time optimization for robust video classification.
\newblock \emph{arXiv preprint arXiv:2302.14309}.

\bibitem[{Yoshimura et~al.(2020)Yoshimura, Hayashi, Takeda, and Watanabe}]{vad_sup}
Takenori Yoshimura, Tomoki Hayashi, Kazuya Takeda, and Shinji Watanabe. 2020.
\newblock End-to-end automatic speech recognition integrated with ctc-based voice activity detection.
\newblock In \emph{ICASSP 2020-2020 IEEE International Conference on Acoustics, Speech and Signal Processing (ICASSP)}, pages 6999--7003. IEEE.

\bibitem[{Zaken et~al.(2021)Zaken, Ravfogel, and Goldberg}]{zaken2021bitfit}
Elad~Ben Zaken, Shauli Ravfogel, and Yoav Goldberg. 2021.
\newblock Bitfit: Simple parameter-efficient fine-tuning for transformer-based masked language-models.
\newblock \emph{arXiv preprint arXiv:2106.10199}.

\bibitem[{Zhao et~al.(2018)Zhao, Sonsaat, Silpachai, Lucic, Chukharev-Hudilainen, Levis, and Gutierrez-Osuna}]{l2arctic}
Guanlong Zhao, Sinem Sonsaat, Alif Silpachai, Ivana Lucic, Evgeny Chukharev-Hudilainen, John Levis, and Ricardo Gutierrez-Osuna. 2018.
\newblock L2-arctic: A non-native english speech corpus.
\newblock In \emph{Interspeech}, pages 2783--2787.

\bibitem[{Zhou et~al.(2023)Zhou, Guo, Jia, Zhang, and Li}]{ods}
Zhi Zhou, Lan-Zhe Guo, Lin-Han Jia, Dingchu Zhang, and Yu-Feng Li. 2023.
\newblock {ODS}: Test-time adaptation in the presence of open-world data shift.
\newblock In \emph{Proceedings of the 40th International Conference on Machine Learning}, volume 202 of \emph{Proceedings of Machine Learning Research}, pages 42574--42588. PMLR.

\end{thebibliography}

\appendix

\section{Experimental Details}
\subsection{Dataset Details}
\label{app:data}
We show the statistics of datasets used in our work in Table \ref{table:app_data} where \# Utt. indicates the total number of utterances. We build our synthetic datasets on LibriSpeech test-other set. For LS-C, we add the Gaussian noises when preparing the data loader and use the amplitudes \{0.005, 0.01, 0.015, 0.02, 0.03\} as level 1-5 severity. For LS-P, we use the AirConditioner\_6, Typing\_2, Babble\_4, Munching\_3, ShuttingDoor\_6, VacuumCleaner\_1, AirportAnnouncements\_2, CopyMachine\_2 wave files from MS-SNSD~\footnote{https://github.com/microsoft/MS-SNSD} as the environmental sounds and synthesize audios with signal-to-noise ratios \{10, 5, 0, -5, -10\} seperately. For L2-Arctic, we use the default splits of 24 non-native speakers with a balanced gender and L1 distribution. For music datasets, we use the default DSing dev and test sets and the full Hansen set (no split).
\begin{table}[htbp]
\begin{center}
\begin{tabular}{lllllccccccccccccccc}
\toprule[1pt]
\multicolumn{5}{l}{Type} & \multicolumn{5}{c}{Datasets} & \multicolumn{5}{c}{\# Utt.} & \multicolumn{5}{c}{Duration}  \\ 
\midrule[0.5pt]
\multicolumn{5}{l}{\multirow{2}{*}{Noise}} & \multicolumn{5}{c}{LS-C} & \multicolumn{5}{c}{14695} & \multicolumn{5}{c}{25.5 h } \\ 
\multicolumn{5}{l}{} & \multicolumn{5}{c}{LS-P} & \multicolumn{5}{c}{117560} & \multicolumn{5}{c}{204 h} \\
\midrule[0.5pt]
\multicolumn{5}{l}{Accent} & \multicolumn{5}{c}{L2-Arctic} & \multicolumn{5}{c}{26867} & \multicolumn{5}{c}{27.1 h}  \\ 
\midrule[0.5pt]
\multicolumn{5}{l}{\multirow{3}{*}{Music}} & \multicolumn{5}{c}{DSing-dev}& \multicolumn{5}{c}{482} & \multicolumn{5}{c}{41 min}  \\ 
\multicolumn{5}{l}{} & \multicolumn{5}{c}{DSing-test}& \multicolumn{5}{c}{480} & \multicolumn{5}{c}{48 min}  \\ 
\multicolumn{5}{l}{} & \multicolumn{5}{c}{Hansen}& \multicolumn{5}{c}{634} & \multicolumn{5}{c}{34 min}  \\ 

\bottomrule[1pt]
\end{tabular}
\caption{Statistics of evaluation datasets.}
\label{table:app_data}
\end{center}
\vspace{-3mm}
\end{table}


\subsection{Implementation Details}
\label{app:implement}
In our experimental evaluations, we mainly employ the acoustic foundation model, Wav2vec2. Specifically, we utilize its Connectionist Temporal Classification (CTC) variants with different model sizes, Wav2vec2 Base and Wav2vec2 Large. We involve the usage of publicly available Wav2vev2 Base~\footnote{https://huggingface.co/facebook/wav2vec2-base-960h} and Wav2vec2 Large~\footnote{https://huggingface.co/facebook/wav2vec2-large-960h-lv60-self} models fine-tuned on speech recognition tasks. The detailed structure of the CTC model is a single fully-connected layer and softmax on top of the foundation model. Given that CTC-based models do not explicitly model silences, we take those with the pseudo label <BLANK> as silent frames and the rest as non-silent frames as per \citep{kurzinger2020ctc,wei2022unsupervised,yang2023blank}. We are interested in those frames carrying important semantic information so we take the blank indicator as an approximation. The advantage is to directly utilize the test-time inference output without additional computation such as a VAD module. Moreover, we found taking the blank symbol as an indicator has already achieved good performance in existing work~\citep{vad_sup} which serves as a good support. We mainly conduct experiments on these two models despite the applicability of our method to other transformer-based architectures of acoustic foundation models. To make a fair comparison with methods employing beam search, we utilize the same 4-gram language model~\footnote{https://huggingface.co/patrickvonplaten/wav2vec2-base-100h-with-lm} as SGEM. Since our test-time setting requires no access to the target text, we use the language model trained on the speech dataset despite the text-domain shift. For the Conformer and Transducer, we employ Conformer-CTC~\footnote{https://catalog.ngc.nvidia.com/orgs/nvidia/teams/nemo/\\models/stt\_en\_conformer\_ctc\_small\_ls} and Conformer-Transducer~\footnote{https://catalog.ngc.nvidia.com/orgs/nvidia/teams/nemo/\\models/stt\_en\_conformer\_transducer\_small}. All speech inputs are sampled or resampled at 16Khz.

We use Pytorch and Huggingface Transformers in our implementation. All experiments are run on a single NVIDIA A5000 GPU (24G). We evaluate the performance of all baselines after adaptation for ten steps. We use the AdamW optimizer as default for all experiments. The weight $\alpha$ of consistency regularization is set to be 0.3. We consider the learning rate in \{2e-4, 5e-4, 8e-4\} for tuning affine parameters of layer normalization and consider the learning rate in \{2e-5, 5e-5\} for tuning feature extractor. Since the TTA setting has no validation set, we follow SUTA and use the hyperparameters obtained from Librispeech test-other set with noise level $\delta = 0.01$ as the default for the experiments. For singing data experiments, we use the hyperparameters obtained from DSing-dev as the default for experiments on DSing-test and Hansen.

\section{More results}

\subsection{More Ablation Study}
\textbf{Strategies for Frame Selection} We proceed to analyze strategies utilized for the selection of speech frames optimized within the CEA framework. We investigate three pseudo-label-based strategies, namely a) selection of non-silent frames (as used in our method), b) selection of silent frames, and c) selection of all frames. The results are detailed in Table \ref{subtable:abl_fil}. The empirical findings reveal that the optimization of silent frames or all frames within CEA yields inferior performance compared to the optimization of non-silent frames. Moreover, it is observed that the degradation is not so substantial, as optimizing silent or all frames may also contribute to enhancing the reliability of noisy frames.

 


\begin{table}[htbp]

\centering
\begin{tabular}{cccccccccccccccccc}
\toprule[1pt]
 \multicolumn{6}{c}{Strategy} & \multicolumn{6}{c}{DSing-dev} & \multicolumn{6}{c}{DSing-test} \\ 
 \midrule[0.1pt]
 \multicolumn{6}{c}{Non-Silent} & \multicolumn{6}{c}{$\textbf{53.5}$} & \multicolumn{6}{c}{$\textbf{50.1}$}  \\ 
\multicolumn{6}{c}{Silent} & \multicolumn{6}{c}{$54.9$} & \multicolumn{6}{c}{$51.7$}   \\ 
\multicolumn{6}{c}{All} & \multicolumn{6}{c}{$54.9$} & \multicolumn{6}{c}{$50.6$}   \\ 
\bottomrule[1pt]
\end{tabular}
\caption{Ablation study of strategies for frame selection. WER (\%) results are reported.}
\label{subtable:abl_fil}
\vspace{-2mm}
\end{table}

\noindent \textbf{Efficacy of STCR on SUTA} To further validate the efficacy of short-term consistency regularization, we did one more ablation study using SUTA + STCR on the DSing-test set, and observed that the proposed SCTR can enhance SUTA with WER decreasing from $51.3$ to $50.9$. However, the performance of SUTA + STCR still lags behind our method CEA + STCR with WER $50.1$, which demonstrates that our proposed CEA also contributes to the final improvement.

\subsection{Comparison of Adaptation Time}
\label{compute_time}
Given the same recognition run-time using the same Wav2vec2 Base, we provide a comparison of the average adaptation time using the DSing-test set on A5000 in Table~\ref{app:compute_time}.

\begin{table}[h]

\centering
\begin{tabular}{cccccccccccc}
\toprule[1pt]
\multicolumn{6}{c}{Method} & \multicolumn{6}{c}{Runtime} \\ 
\midrule[0.5pt]
\multicolumn{6}{c}{Tent} & \multicolumn{6}{c}{$0.328$} \\ 
\multicolumn{6}{c}{SAR}  & \multicolumn{6}{c}{$0.733$} \\ 
\multicolumn{6}{c}{TeCo} & \multicolumn{6}{c}{$0.401$} \\ 
\multicolumn{6}{c}{SUTA} & \multicolumn{6}{c}{$0.483$}  \\  
\multicolumn{6}{c}{Ours} & \multicolumn{6}{c}{$0.879$}  \\ 
\bottomrule[1pt]
\end{tabular}
\caption{Comparison of adaptation time using Wav2Vec2 Base.}
\label{app:compute_time}

\end{table}

\subsection{Results on CHiME3}
We have conducted additional experiments on CHiME3 using Wav2vec2 Base, and the results, shown in Table~\ref{app:chime}, demonstrate that our method outperforms other baselines.
\begin{table}[h]

\centering
\begin{tabular}{cccccccccccc}
\toprule[1pt]
\multicolumn{6}{c}{Method} & \multicolumn{6}{c}{WER} \\ 
\midrule[0.5pt]
\multicolumn{6}{c}{Source} & \multicolumn{6}{c}{$31.2$} \\ 
\multicolumn{6}{c}{Tent} & \multicolumn{6}{c}{$28.0$} \\ 
\multicolumn{6}{c}{SAR}  & \multicolumn{6}{c}{$29.0$} \\ 
\multicolumn{6}{c}{TeCo} & \multicolumn{6}{c}{$28.0$} \\ 
\multicolumn{6}{c}{SUTA} & \multicolumn{6}{c}{$25.0$}  \\  
\multicolumn{6}{c}{Ours} & \multicolumn{6}{c}{$24.5$}  \\ 
\bottomrule[1pt]
\end{tabular}
\caption{WER (\%) results on CHiME3 using Wav2vec2 Base with greedy decoding.}
\label{app:chime}

\end{table}

\subsection{Results on More Acoustic Foundation Models}
\label{app: more_models}


\begin{table*}[htbp]
  \centering

    \begin{tabular}{cccccccc}
    \toprule
          & Size & Level 1 & Level 2 & Level 3 & Level 4 & Level 5 & Avg \\
    \midrule
    \multicolumn{1}{l}{Wav2vec2} &       &       &       &       &       &  \\
    \midrule
    \multirow{2}{*}{Source} & Base & 13.9  & 24.4  & 39.5  & 54.5  & 75.7  & 41.6 \\
    & Large & 5.0   & 8.1   & 14.6  & 24.9  & 46.9  & 19.9 \\
    \midrule
    \multirow{2}{*}{Ours}  & Base & 10.7  & 16.2  & 24.0  & 34.1  & 56.5  & 28.3 \\
    & Large & 4.3   & 6.1   & 9.7   & 15.1  & 31.1  & 13.3\\
    \midrule
    \multirow{2}{*}{WERR (\%)}  & Base & 23.0 & 33.6 & 39.2 & 37.4 & 25.4 & 31.7 \\
    & Large & 14.0 & 24.7 & 33.6 & 39.4 & 33.7 & 29.1 \\
    \midrule
    \multicolumn{1}{l}{Hubert} &       &       &       &       &       &  \\
    \midrule
    \multirow{2}{*}{Source} & Base & 26.1  & 32.7  & 40.6  & 49.0  & 63.4  & 42.4 \\
    & Large & 5.0   & 6.4   & 8.9   & 12.8  & 24.3  & 11.5 \\
    \midrule
    \multirow{2}{*}{Ours}  & Base & 19.3  & 23.7  & 28.9  & 35.0  & 47.5  & 30.9 \\
    & Large & 4.3   & 5.2   & 6.9   & 9.1   & 16.1  & 8.3 \\
    \midrule
    \multirow{2}{*}{WERR (\%)}  & Base & 26.1 & 27.5 & 28.8 & 28.6 & 25.1 & 27.2 \\
    & Large & 14.0 & 18.8 & 22.5 & 28.9 & 33.7 & 23.6 \\
    \midrule
    \multicolumn{1}{l}{WavLM} &       &       &       &       &       &  \\
    \midrule
    \multirow{2}{*}{Source} & Base & 24.1  & 35.9  & 48.2  & 59.8  & 76.7  & 48.9 \\
    & Large & 14.4  & 17.5  & 21.5  & 26.1  & 36.1  & 23.1 \\
    \midrule
    \multirow{2}{*}{Ours}  & Base & 15.1  & 19.8  & 25.9  & 32.8  & 47.6  & 28.2 \\
    & Large & 10.7  & 12.4  & 14.5  & 17.1  & 23.9  & 15.7 \\
    \midrule
    \multirow{2}{*}{WERR (\%)}  & Base & 37.3 & 44.8 & 46.3 & 45.2 & 37.9 & 42.3  \\
    & Large & 25.7 & 29.1 & 32.6 & 34.5 & 33.8 & 31.1 \\
    \bottomrule
    \end{tabular}%
  \caption{WER (\%) results on LS-C over five severity levels of Gaussian noises using both base and large models of Wav2vec2, Hubert, WavLM with greedy decoding. WERR stands for word error rate reduction.}
  \label{tab:diverse_models}%
\end{table*}%

\begin{table*}[htbp]
  \centering
    \begin{tabular}{ccccccc}
    \toprule
    & \multicolumn{2}{c}{Wav2vec2} & \multicolumn{2}{c}{Hubert} & \multicolumn{2}{c}{WavLM} \\
    \midrule
    & Base  & Large  & Base  & Large  & Base  & Large \\
    \midrule
    Source & 60.1  & 38.8  & 71.5  & 43.9  & 76.1  & 66.2 \\
    Ours  & 50.1  & 31.2  & 62.4  & 32.4  & 59.6  & 51.1 \\
    WERR (\%) & 16.6 & 19.6 & 12.7 & 26.2 & 21.7 & 22.8 \\
    \bottomrule
    \end{tabular}%
    \caption{WER (\%) results on DSing-test using both base and large models of Wav2vec2, Hubert, WavLM with greedy decoding. WERR stands for word error rate reduction.}
  \label{tab:models_dsing}%
\end{table*}%

In an extension of the main experiments, we delved into the adaptation performance across diverse acoustic foundation models. Specifically, our additional experiments utilize various models including, Hubert-Base~\footnote{https://huggingface.co/danieleV9H/hubert-base-libri-clean-ft100h}, Hubert-Large~\footnote{https://huggingface.co/facebook/hubert-large-ls960-ft}, WavLM-Base~\footnote{https://huggingface.co/patrickvonplaten/wavlm-libri-clean-100h-base-plus}, and WavLM-Large~\footnote{https://huggingface.co/patrickvonplaten/wavlm-libri-clean-100h-large} from Huggingface. These experiments are conducted to assess the adaptation performance ain relation to different model sizes, and training data sources. The outcomes on the LS-C and DSing-test datasets are reported in Table \ref{tab:diverse_models} and Table \ref{tab:models_dsing} respectively. We employ the word error rate reduction (WERR) to measure the relative improvement brought by our adaptation method. We summarize the findings as follows:

\textbf{Model Sizes}. A comparative analysis is conducted between the base and large versions of each model. The findings reveal that large models consistently surpass base models. Furthermore, our proposed approach uniformly improves both base and large models. A notable observation is that our method elicits a greater average improvement in base models compared to large models within the LS-C dataset. This trend is particularly pronounced under lower noise levels ranging from 1 to 3. In contrast, within the DSing-test set, the enhancement for large models is more significant than for base models. The phenomenon may be attributed to the fact that large models already exhibit commendable performance under minor corruptions, even without adaptation, thus providing limited scope for further improvement. However, in scenarios involving significant shifts, the expansive parameterization of large models facilitates more effective adaptation, whereas base models face challenges.

\textbf{Training Data Sources}. A comparative evaluation of models trained with different datasets, including Wav2vec2-Large trained with 960h LibriSpeech set, Hubert-Large trained with 960h LibriSpeech set, and WavLM-Large trained with 100h LibriSpeech clean set, indicates that the larger-size data set establish a stronger foundation for test-time adaptation. A similar inference can be drawn when comparing Wav2vec2-Base trained with 960h LibriSpeech set, Hubert-Base trained with 100h LibriSpeech clean set, and WavLM-Base trained with 100h LibriSpeech clean set.

In summary, our proposed unsupervised TTA method demonstrates a considerable benefit across diverse acoustic foundation models, reflecting substantial improvements for different model sizes and training data sources. 



\subsection{Analysis on Large Vocabulary Size}
Our proposed method can be generalizable to models with large vocabulary sizes. Theoretically, the maximum entropy for non-silent frames is expected to increase due to the larger number of classes. Practically, this might also depend on the test input and models. To analyze the entropy distribution for non-silent and silent frames, we conduct an additional experiment using the Conformer-CTC model with BPE tokenization, which has a larger vocabulary size than the one of the Wav2vec2 model. We observed an increase in entropy for non-silent frames from 59.4\% to 70.0\%, as illustrated in Table \ref{subtable:vocab_size}.

\begin{table}[htbp]

\centering
\begin{tabular}{cccccccccccccccccc}
\toprule[1pt]
 \multicolumn{6}{c}{} & \multicolumn{6}{c}{Wav2vec2 Base} & \multicolumn{6}{c}{Conformer-CTC} \\ 
 \midrule[0.1pt]
 \multicolumn{6}{c}{n-sil-h} & \multicolumn{6}{c}{$0.594$} & \multicolumn{6}{c}{$0.700$}  \\ 
\multicolumn{6}{c}{n-sil-l} & \multicolumn{6}{c}{$0.406$} & \multicolumn{6}{c}{$0.300$}   \\ 
\multicolumn{6}{c}{sil-h} & \multicolumn{6}{c}{$0.362$} & \multicolumn{6}{c}{$0.497$}   \\ 
\multicolumn{6}{c}{sil-l} & \multicolumn{6}{c}{$0.638$} & \multicolumn{6}{c}{$0.503$}   \\ 
\bottomrule[1pt]
\end{tabular}
\caption{Entropy Distribution at Step 0 for models with different vocabulary sizes. "non-sil" and "sil" refer to non-silent and silent frames, respectively. "h / l" indicates frames with high or low entropy.}
\label{subtable:vocab_size}
\end{table}

\subsection{Connection with Existing Frozen Model Adaptation}
Our TTA-based method also exhibits parameter efficiency. It is essential to emphasize that our approach does not introduce additional layers of normalization. Instead, we adapt the affine parameters (the scale $\gamma$ and the shift $\beta$) of the existing layer normalization from the pre-training phase, which means no new trainable parameters are introduced. It is noteworthy to highlight the difference between our method and existing frozen model adaptation methods, such as P-tuning, LoRA, and Adapter. Unlike these techniques, our method conducts source-free unsupervised adaptation using a single utterance. Furthermore, our primary objective of adaptation is to address open-world acoustic data shifts, rather than task adaptation.

\subsection{Results on Different Parameterizations}
\begin{table}

\centering
\begin{tabular}{ccccccccccccccc}
\toprule[1pt]
 \multicolumn{3}{c}{\multirow{2}{*}{Type}} & \multicolumn{6}{c}{\multirow{1}{*}{Base}} & \multicolumn{6}{c}{\multirow{1}{*}{Large}} \\ 
 \cline{4-15}
 \multicolumn{3}{c}{} & \multicolumn{3}{c}{\multirow{1}{*}{WER}} &\multicolumn{3}{c}{\multirow{1}{*}{Params}} & \multicolumn{3}{c}{\multirow{1}{*}{WER}} & \multicolumn{3}{c}{\multirow{1}{*}{Params}} \\ 
 \midrule[0.5pt]
 \multicolumn{3}{c}{Bias-Only} & \multicolumn{3}{c}{$52.5$} &\multicolumn{3}{c}{$0.10$M} & \multicolumn{3}{c}{$31.8$} & \multicolumn{3}{c}{$0.28$M} \\ 
 \multicolumn{3}{c}{LNs} & \multicolumn{3}{c}{$52.4$} &\multicolumn{3}{c}{$0.04$M} & \multicolumn{3}{c}{$31.4$} & \multicolumn{3}{c}{$0.11$M} \\ 
 \multicolumn{3}{c}{FE+LNs} & \multicolumn{3}{c}{$\textbf{50.1}$} &\multicolumn{3}{c}{$4.63$M} & \multicolumn{3}{c}{$\textbf{31.2}$} & \multicolumn{3}{c}{$4.84$M} \\ 
 \midrule[0.5pt]
 \multicolumn{3}{c}{Full} & \multicolumn{3}{c}{$51.2$} &\multicolumn{3}{c}{$89.7$M} & \multicolumn{3}{c}{$31.9$} & \multicolumn{3}{c}{$307$M} \\ 

\bottomrule[1pt]
\end{tabular}
\caption{Results with different parameterizations on DSing-test using Wav2vec2 Base and Large models. We consider (1) Bias-Only: all bias terms, (2) LNs: all scale and shift terms of Layer Normalization, 3) FE+LNs: parameters of the feature extractor and all scale and shift terms of Layer Normalization, and (4) Full: all parameters. Word Error Rate (\%) and the number of parameters (Params) are reported. }
\label{subtable:abl_params}

\vspace{-3mm}
\end{table}

In order to further evaluate the effectiveness of our proposed method across diverse parameterizations, we conduct additional experiments on the DSing-test set using Wav2vec2 Base and Large models. Specifically, we explore four distinct parameterization schemes and compute their corresponding number of parameters: (1) \textbf{Bias-Only} refers to fine-tuning only bias terms as per \citet{zaken2021bitfit}. (2) \textbf{LNs} encompasses the adjustment of all scale and shift terms associated with layer normalization. (3) \textbf{FE+LNs} involves the parameters of the feature extractor in addition to all scale and shift terms of layer normalization. (4) \textbf{Full} entails the fine-tuning of all parameters within the model. It is important to note that all other experimental settings except for parameterization have remained consistent. The experimental results are presented in Table \ref{subtable:abl_params}. Our findings reveal that our method exhibits compatibility with different parameterizations, yielding comparable performances. Among these parameterizations, LNs demonstrate the smallest number of parameters adjusted, thereby illustrating the parameter efficiency of our method.

\subsection{Full Results for LS-P}
\label{app:ls-p}
We present the full WER results for eight environmental sounds of five severity levels in Table~\ref{subtable:aircon} - \ref{subtable:babble}. The first row denotes signal-to-noise ratios.

\subsection{Full Results for L2-Arctic}
\label{app:l2}
We present the full speaker-level WER results for each L1 in Table~\ref{subtable:arabic} - \ref{subtable:vietnamese}. The first row denotes the speaker ID. The details of the speaker ID can be found in the L2-Arctic~\footnote{https://psi.engr.tamu.edu/l2-arctic-corpus/}.

\newpage
\begin{table*}
\centering
\begin{minipage}{0.47\textwidth}
\centering
\begin{tabular}{cccccc}
    \toprule
          & 10    & 5     & 0     & -5    & -10 \\
    \midrule
    Source & 28.1  & 43.9  & 65.0  & 83.4  & 94.2 \\
    Tent  & 22.6  & 36.1  & 56.6  & 77.9  & 91.4 \\
    SAR   & 24.5  & 39.1  & 59.9  & 79.9  & 92.1 \\
    TeCo  & 22.5  & 36.2  & 56.6  & 77.9  & 91.3 \\
    SUTA  & 17.7  & 26.1  & 41.2  & 62.7  & 82.7 \\
    Ours  & \textbf{17.5}  & \textbf{25.6}  & \textbf{40.6}  & \textbf{61.6}  & \textbf{82.2} \\
    \bottomrule
    \end{tabular}%
\caption{Air Conditioner.}
\label{subtable:aircon}
\end{minipage}%
\hfill
\begin{minipage}{0.47\textwidth}
\centering
    \begin{tabular}{cccccc}
    \toprule
          & 10    & 5     & 0     & -5    & -10 \\
    \midrule
    Source & 26.2  & 34.0  & 44.4  & 56.4  & 69.0 \\
    Tent  & 21.0  & 27.9  & 37.0  & 49.2  & 63.0 \\
    SAR   & 23.0  & 30.3  & 39.7  & 52.1  & 65.3 \\
    TeCo  & 21.0  & 27.8  & 37.0  & 49.1  & 63.0 \\
    SUTA  & 17.9  & 23.3  & 30.4  & 41.0  & 53.4 \\
    Ours  & \textbf{17.5}  & \textbf{22.8}  & \textbf{29.9}  & \textbf{40.4}  & \textbf{52.6} \\
    \bottomrule
    \end{tabular}%
\caption{Typing.}
\label{subtable:typing}
\end{minipage}

\end{table*}

\begin{table*}
\centering
\begin{minipage}{0.47\textwidth}
\centering
    \begin{tabular}{cccccc}
    \toprule
          & 10    & 5     & 0     & -5    & -10 \\
    \midrule
    Source & 50.4  & 62.8  & 74.6  & 83.8  & 90.1 \\
    Tent  & 44.8  & 57.6  & 71.1  & 82.7  & 90.5 \\
    SAR   & 47.3  & 57.8  & 72.1  & 82.5  & 89.6 \\
    TeCo  & 44.8  & 57.6  & 71.1  & 82.7  & 90.5 \\
    SUTA  & 39.7  & 51.9  & 64.4  & 76.4  & \textbf{85.2} \\
    Ours  & \textbf{39.3}  & \textbf{51.5}  & \textbf{64.1}  & \textbf{76.3}  & 85.3 \\
    \bottomrule
    \end{tabular}%
\caption{Munching.}
\label{subtable:munching}
\end{minipage}%
\hfill
\begin{minipage}{0.47\textwidth}
\centering
    \begin{tabular}{cccccc}
    \toprule
          & 10    & 5     & 0     & -5    & -10 \\
    \midrule
    Source & 19.2  & 23.6  & 29.7  & 37.0  & 45.0 \\
    Tent  & 16.4  & 20.5  & 26.0  & 33.0  & 41.5 \\
    SAR   & 17.7  & 22.0  & 27.7  & 35.0  & 42.7 \\
    TeCo  & 16.3  & 20.5  & 26.0  & 32.9  & 41.5 \\
    SUTA  & 14.9  & 18.5  & 23.6  & 29.9  & 37.7 \\
    Ours  & \textbf{14.8}  & \textbf{18.3}  & \textbf{23.4}  & \textbf{29.7}  & \textbf{37.4} \\
    \bottomrule
    \end{tabular}%
\caption{Shutting Door.}
\label{subtable:shuttingdoor}
\end{minipage}

\end{table*}

\begin{table*}
\centering
\begin{minipage}{0.47\textwidth}
\centering
\begin{tabular}{cccccc}
    \toprule
          & 10    & 5     & 0     & -5    & -10 \\
    \midrule
    Source & 57.8  & 76.6  & 91.5  & 98.2  & 99.9 \\
    Tent  & 49.7  & 69.2  & 87.2  & 97.0  & 99.6 \\
    SAR   & 52.6  & 72.7  & 88.5  & 96.9  & 99.8 \\
    TeCo  & 49.7  & 69.2  & 87.2  & 96.9  & 99.6 \\
    SUTA  & 39.8  & 56.7  & 76.6  & 93.2  & \textbf{98.6} \\
    Ours  & \textbf{39.3}  & \textbf{56.0}  & \textbf{76.0}  & \textbf{93.0}  & \textbf{98.6} \\
    \bottomrule
    \end{tabular}%
\caption{Vacuum Cleaner.}
\label{subtable:vacuumclean}
\end{minipage}%
\hfill
\begin{minipage}{0.47\textwidth}
\centering
    \begin{tabular}{cccccc}
    \toprule
          & 10    & 5     & 0     & -5    & -10 \\
    \midrule
    Source & 40.9  & 54.3  & 66.3  & 75.8  & 83.4 \\
    Tent  & 36.1  & 49.3  & 62.8  & 73.7  & 82.4 \\
    SAR   & 38.2  & 51.0  & 64.0  & 74.3  & 82.2 \\
    TeCo  & 36.1  & 49.2  & 62.8  & 73.7  & 82.3 \\
    SUTA  & \textbf{31.2}  & 43.8  & 58.3  & \textbf{70.4}  & \textbf{79.3} \\
    Ours  & \textbf{31.2}  & \textbf{43.7}  & \textbf{58.1}  & 70.5  & 79.7 \\
    \bottomrule
    \end{tabular}%
\caption{Airpoint Announcements.}
\label{subtable:airpoitann}
\end{minipage}

\end{table*}

\begin{table*}
\centering
\begin{minipage}{0.47\textwidth}
\centering
    \begin{tabular}{cccccc}
    \toprule
          & 10    & 5     & 0     & -5    & -10 \\
    \midrule
    Source & 49.8  & 63.5  & 76.6  & 86.9  & 93.5 \\
    Tent  & 44.4  & 58.9  & 74.2  & 86.3  & 93.7 \\
    SAR   & 46.6  & 60.7  & 74.8  & 86.2  & 93.2 \\
    TeCo  & 44.4  & 58.8  & 74.2  & 86.2  & 93.7 \\
    SUTA  & 39.3  & 52.7  & 67.4  & \textbf{80.8}  & \textbf{89.7} \\
    Ours  & \textbf{38.9}  & \textbf{52.3}  & \textbf{67.3}  & 81.0  & 89.8 \\
    \bottomrule
    \end{tabular}%
\caption{Copy Machine.}
\label{subtable:copymachine}
\end{minipage}%
\hfill
\begin{minipage}{0.5\textwidth}
\centering
    \begin{tabular}{cccccc}
    \toprule
          & 10    & 5     & 0     & -5    & -10 \\
    \midrule
    Source & 66.6  & 81.6  & 94.7  & 104.3 & 111.2 \\
    Tent  & 62.0  & 77.8  & 92.0  & 102.2 & 109.4 \\
    SAR   & 62.8  & 77.7  & 90.5  & 102.1 & \textbf{106.9} \\
    TeCo  & 61.9  & 77.8  & 91.9  & 102.2 & 109.4 \\
    SUTA  & \textbf{55.5}  & \textbf{73.0}  & \textbf{88.6}  & \textbf{101.1} & 109.2 \\
    Ours  & \textbf{55.5}  & \textbf{73.0}  & 89.1  & 102.0 & 110.3 \\
    \bottomrule
    \end{tabular}%
\caption{Babble.}
\label{subtable:babble}

\end{minipage}

\end{table*}

\clearpage
\begin{table*}
\centering
\begin{minipage}{0.47\textwidth}
\centering
    \begin{tabular}{ccccc}
    \toprule
          & ABA   & SKA   & YBAA  & ZHAA \\
    \midrule
    Source & 21.0  & 32.5  & 16.7  & 17.3 \\
    Tent  & 18.4  & 28.4  & 14.5  & 14.4 \\
    SAR   & 19.4  & 30.3  & 15.7  & 15.3 \\
    TeCo  & 18.4  & 28.4  & 14.5  & 14.4 \\
    SUTA  & 17.8  & 27.2  & 13.7  & 14.0 \\
    Ours  & \textbf{17.7}  & \textbf{26.8}  & \textbf{13.5}  & \textbf{13.9} \\
    \bottomrule
    \end{tabular}%
\caption{Arabic.}
\label{subtable:arabic}
\end{minipage}%
\hfill
\begin{minipage}{0.47\textwidth}
\centering
\begin{tabular}{ccccc}
    \toprule
          & BWC   & LXC   & NCC   & TXHC \\
    \midrule
    Source & 28.5  & 33.5  & 26.9  & 21.1 \\
    Tent  & 24.1  & 29.2  & 22.8  & 18.1 \\
    SAR   & 26.3  & 30.9  & 25.0  & 19.5 \\
    TeCo  & 24.1  & 29.3  & 22.9  & 18.0 \\
    SUTA  & 23.3  & \textbf{27.6}  & 21.5  & 17.4 \\
    Ours  & \textbf{23.0}  & 27.7  & \textbf{21.3}  & \textbf{17.3} \\
    \bottomrule
    \end{tabular}%
\caption{Mandarin.}
\label{subtable:mandarin}
\end{minipage}

\end{table*}

\begin{table*}
\centering
\begin{minipage}{0.47\textwidth}
\centering
    \begin{tabular}{ccccc}
    \toprule
          & ASI   & RRBI  & SVBI  & TNI \\
    \midrule
    Source & 14.3  & 15.7  & 19.8  & 18.6 \\
    Tent  & 11.7  & 12.9  & 15.7  & 15.6 \\
    SAR   & 12.7  & 14.0  & 17.6  & 16.7 \\
    TeCo  & 11.7  & 13.0  & 15.8  & 15.6 \\
    SUTA  & \textbf{11.3}  & 12.5  & 14.3  & 14.9 \\
    Ours  & \textbf{11.3}  & \textbf{12.2}  & \textbf{14.3}  & \textbf{14.8} \\
    \bottomrule
    \end{tabular}%
\caption{Hindi.}
\label{subtable:hindi}
\end{minipage}%
\hfill
\begin{minipage}{0.47\textwidth}
\centering
    \begin{tabular}{ccccc}
    \toprule
          & HJK   & HKK   & YDCK  & YKWK \\
    \midrule
    Source & 11.8  & 23.3  & 17.2  & 17.0 \\
    Tent  & 9.7   & 20.8  & 15.0  & 14.5 \\
    SAR   & 10.9  & 21.7  & 15.8  & 15.5 \\
    TeCo  & 9.8   & 20.8  & 15.0  & 14.5 \\
    SUTA  & \textbf{9.5}   & 19.8  & 14.2  & 13.8 \\
    Ours  & \textbf{9.5}   & \textbf{19.7}  & \textbf{13.9}  & \textbf{13.7} \\
    \bottomrule
    \end{tabular}%
\caption{Korean.}
\label{subtable:korean}
\end{minipage}

\end{table*}

\begin{table*}
\centering
\begin{minipage}{0.49\textwidth}
\centering
    \begin{tabular}{ccccc}
    \toprule
          & EBVS  & ERMS  & MBMPS & NJS \\
    \midrule
    Source & 35.7  & 24.2  & 14.1  & 14.6 \\
    Tent  & 31.7  & 20.0  & 12.7  & 12.4 \\
    SAR   & 33.5  & 21.7  & 13.4  & 13.2 \\
    TeCo  & 31.7  & 20.0  & 12.7  & 12.4 \\
    SUTA  & 29.7  & 18.7  & \textbf{12.3}  & \textbf{12.1} \\
    Ours  & \textbf{29.5}  & \textbf{18.5}  & \textbf{12.3}  & \textbf{12.1} \\
    \bottomrule
    \end{tabular}%
\caption{Spanish.}
\label{subtable:spanish}
\end{minipage}%
\hfill
\begin{minipage}{0.47\textwidth}
\centering
    \begin{tabular}{ccccc}
    \toprule
          & HQTV  & PNV   & THV   & TLV \\
    \midrule
    Source & 41.6  & 18.5  & 38.1  & 41.1 \\
    Tent  & 38.0  & 16.4  & 34.4  & 38.1 \\
    SAR   & 40.3  & 17.6  & 36.2  & 39.4 \\
    TeCo  & 38.0  & 16.4  & 34.4  & 38.0 \\
    SUTA  & 36.5  & \textbf{15.5}  & 33.2  & \textbf{36.8} \\
    Ours  & \textbf{36.3}  & \textbf{15.5}  & \textbf{32.9}  & \textbf{36.8} \\
    \bottomrule
    \end{tabular}%
\caption{Vietnamese.}
\label{subtable:vietnamese}
\end{minipage}

\end{table*}

\end{document}